\documentclass[11pt]{article}
\usepackage{amsmath, amssymb, amscd, amsthm, amsfonts}
\usepackage{graphicx}
\usepackage[utf8]{inputenc}
\usepackage[T1]{fontenc}
\usepackage{color}
\usepackage{hyperref}
\usepackage{relsize}
\usepackage{graphics} 
\usepackage{epsfig}
\usepackage{mathptmx} 
\usepackage{algorithm}

\usepackage{algpseudocode}
\usepackage{lmodern} 
\usepackage{diagbox}
\usepackage{mathtools}
\DeclareMathOperator{\sinc}{sinc}

\newtheorem{lemma}{Lemma}
\newtheorem{theorem}{Theorem}
\newtheorem{proposition}{Proposition}
\newtheorem{remark}{Remark}

\newtheorem{assumption}{Assumption}
\newtheorem{definition}{Definition}

\theoremstyle{Proof}

\title{Contribution to the initialization of linear non-commensurate fractional-order systems for the joint estimation of parameters and fractional differentiation orders
}
\author{Mohamed A. Bahloul$^{1,2}$, Zehor Belkhatir$^{3}$ and Taous-Meriem Laleg Kirati$^{2,4}$\\
$^{1}$ College of Engineering, Electrical Engineering Department at Alfaisal University,\\ Riyadh 11533, Saudi Arabia.
{\tt\small E-mail: {mbahloul$@$alfaisal.edu}}\\
$^{2}$ Electrical and Computer Engineering Department, KAUST, Saudi Arabia.\\ 
{\tt\small E-mail: mohamad.bahloul@kaust.edu.sa, taousmeriem.laleg@kaust.edu.sa}\\
$^{3}$ School of Engineering and Sustainable Development, De Montfort University,\\ United Kingdom, {\tt\small E-mail: zehor.belkhatir@dmu.ac.uk}\\
$^{4}$ National Institute for Research in Digital Science and Technology, Paris-Saclay, France.}

\begin{document}
\maketitle
\begin{abstract}
\textbf{It has been recognized that using time-varying initialization functions to solve the initial value problem of fractional-order systems (FOS) is both complex and essential in defining the dynamical behavior of the states of FOSs. In this paper, we investigate the use of the initialization functions for the purpose of estimating unknown parameters of linear non-commensurate FOSs. In particular, we propose a novel "pre-initial" process that describes the dynamic characteristic of FOSs before the initial state and consists of designing an appropriate time-varying initialization function that ensures accurate convergence of the estimates of the unknown parameters. To do so, we propose an estimation technique that consists of two steps:
(i) to design of practical initialization function that is output-dependent and which is employed; (ii) to solve the joint estimation problem of both parameters and fractional differentiation orders (FDOs). A convergence proof has been presented. The performance of the proposed method is illustrated through different numerical examples. Potential applications of the algorithm to joint estimation of parameters and FDOs of the fractional arterial Windkessel and neurovascular models are also presented using both synthetic and real data. The added value of the proposed "pre-initial" process to solve the studied estimation problem is shown through different simulation tests that investigate the sensitivity of estimation results using different time-varying initialization functions.}
\end{abstract}

\section{Introduction}\label{section-introduction}
\noindent Over the past 325 years, fractional calculus (FC) has attracted the attention of mathematicians and engineers working in various fields of science and engineering \cite{gorenflo1997fractional}. FC started to be used as a powerful tool to describe and explore complicated dynamical systems of real-world applications, e.g., biology \cite{magin2010fractional}, control \cite{matuvsuu2011application}, economic \cite{tarasov2019history}. Indeed FC can describe systems with high-order dynamics and complex nonlinear aspects employing fewer coefficients than the integer-order calculus. The arbitrary order of the derivatives provides an extra degree of freedom to concisely and precisely entail the hidden dynamics having a different origin of memory effect. A distinctive feature of fractional-order calculus is that contrary to integer-order calculus, the fractional-order one accounts for the system's memory. In fact, the fractional-order derivative depends not only on the local conditions of the evaluated time but also on the entire history of the function. This peculiarity is usually valuable when the studied system holds a long-term "memory," and any assessed point depends on the past values of the function \cite{podlubny1994fractional}. Accordingly, one essentially-crucial prerequisite to ensuring the fractional-order theory functioning is the adequate initialization and the proper incorporation of the history of the system \cite{lorenzo2008initialization}.

\noindent Generally, the initialization of fractional-order systems is considered complicated and challenging. This dilemma surfaced notably in the case of the non-zero initial value. In fact, it is challenging to design appropriate initialization functions that guarantee the acquiring of exact states of the system while accounting for its history, \cite{bahloul2022initialization}. The pre-initialization process and the initialization functions of non-zero initial value-based fractional-order systems remain an open and controversial question. Various effective methods have been developed to analyze the characteristic of fractional order derivatives where initial values are insufficient \cite{frej2020fractional,ye2015initial,lorenzo2001initialization}. Lorenzo et al., \cite{lorenzo2007initialization} was the first to demonstrate that time-varying functions are more suitable for the initialization of fractional-order systems rather than constant ones. The time-varying initialization has a profound effect on the standard definitions of fractional-order derivative and integral. Different time-varying initialization functions may lead to the same initial value from where the fractional-order operator starts; however, as the system's dynamic is related to the pre-initial process, different initialization functions result in different responses. This fact is known as the aberration phenomenon \cite{du2016estimation}. Accordingly, designing the proper initialization function to acquire the desired convergence of the fractional differential system is deemed very problematic.

\noindent For this reason, in this report, we introduce a novel pre-initialization process that warrants a fast and precise convergence of the joint estimation of the parameters and differentiation orders of the fractional differential system (FOS). The key hypothesis in the design is to consider an output-dependent initialization function when estimating the unknowns. This will reduce the infinite-dimensional space of initialization functions into a finite parametric space where the remaining degree of freedom is about the length of the history function to consider. In addition, we proposed a two-stage algorithm: while the first stage solves a system of linear equations to estimate the parameters of the FOS, the second stage uses the iterative first-order Newton's method \cite{liu2013fractional, belkhatir2015fractional, belkhatir2018parameters, bahloul2021finite} to estimate the fractional differentiation orders. Our contributions are as follows: (i) we consider a time-varying function to initialize the FOS; (ii) we design a pre-initialization process based on the output signal; and (iii) we solve a simple linear equation system to estimate the parameters, and iteratively we apply Newton's method to estimate the fractional differentiation orders.

The performance of the proposed method is illustrated through different numerical examples. Additionally, potential applications of the algorithm are presented, which consists of estimating parameters and fractional differentiation orders of a fractional-order arterial Windkessel and neurovascular models. To the best to the author's knowledge, this is the first study that accounts for the initialization function as part of the parameters estimation problem for fractional differential systems. 
\section{Notations}
For a clear perception of the report, in this section, we present the adopted notation and the basic definitions of the fractional-order integral and derivative. 
Through the following we consider a smooth function $f(t)$ such that $f(t)$ is zero for $t\leq t_{abs}$ and $f(t)$ is $f_{in}(t)$ for $t_{abs}\leq t \leq t_{in}$. 
 \begin{itemize}
    \item $_{t_{in}}D_t^{\alpha} f(t)$ denotes the 'initialized' $\alpha^{th}$ order differ-integration of $f (t)$ from start point $t_0$ to $t$.
    \item $ _{t_{in}}d_t^{\alpha} f(t)$ represents the 'non-initialized' generalized (or fractional) $\alpha^{th}$ order differ-integration of $f(t)$. 
    It is equivalent to shifting the origin of function $f(t)$ at the start of the point from where differ-integration starts.
   \begin{equation}
      \dfrac{d^\alpha f(t)}{[d(t-{t_{in}})^\alpha]}\equiv \ \ _{{t_{in}}}d_t^{\alpha} f(t).
      \label{pre1}
    \end{equation}
    \item $\Psi$ denotes the time-varying initialization function or history function. It is also known as the complementary function \cite{das2011initialized}.
The expression between initialized differ-integral and non-initialized ones is:
      \begin{equation}
     _{t_{in}}D_t^{\alpha} f(t)=_{t_{in}}d_t^{\alpha} f(t)+ \Psi(f,\alpha,t_{abs},t_{in},t)
     \label{pre2}
     \end{equation}
\end{itemize}
%%%%%%%%%%%%%%%%%%%%%%%%%%%%%%%%%%%%%%%%%%%%%%%%%%%%%%%%%%%%%%%%%%%%%%%%%%%%%%%%%%%
\subsection{Definitions}
There are many kinds of definitions for fractional-order differ-integration. Here we present the more commonly known ones, namely the \textit {Riemann-Liouville, Caputo and  Grunwald-Letnikov} definitions. A detailed note on the different definitions might be found in this reference, \cite{lorenzo2007initialization,das2011initialized}.
\begin{definition}
\label{def_RL}
  The \textit{Riemann-Liouville (R-L)} definition of fractional-order integration is drawn as:
\begin{equation}
 \label{RL_I}
  _{t_{in}}^{RL}d_t^{-\alpha}f\left ( t \right )=\frac{1}{\Gamma \left ( \alpha  \right )}\int_{t_{in}}^{t}\frac{f\left ( \tau  \right )}{\left ( t-\tau  \right )^{\alpha }}d\tau,
\end{equation}
here $t_{abs}$ is noted as the terminal point as well. $0<\alpha<1$ and  $\Gamma (x)$ corresponds to so-called \textit{Gamma-function}, written as $\Gamma (x)=\int_{0}^{\infty}e^{-u}u^{x-1}du.$
\end{definition}
\begin{definition}
\label{def_RL_derivative}
 The \textit{R-L} definition of fractional-order derivative is based on the above definition (\ref{RL_I}) and the standard integer-order derivative:
\begin{equation}
  \label{RL_D}
  _{t_{in}}^{RL}d_t^{\alpha}f\left ( t \right )=\frac{\mathrm{d} }{\mathrm{d} t}\left[_{t_{in}}d_t^{-(1-\alpha)}f\left ( t \right )\right].
 \end{equation}
\end{definition}
\begin{definition}
\label{def_caputo}
  The Caputo definition of fractional-order differentiation takes the integer-order differentiation of the function first and then takes a fractional-order integration:
 \begin{equation}
 \label{C_ID}
   _{t_{in}}^Cd_t^{\alpha}f\left ( t \right )=\dfrac{1}{\Gamma \left ( n-\alpha  \right )}\int_{ {t_{in}}}^{t}\left ( 1-\tau  \right )^{-\alpha-1+n } \dfrac{\mathrm{d^n} }{\mathrm{d} t^n}f\left ( \tau  \right )d\tau.
\end{equation}
\end{definition}
\begin{definition}
\label{def_GL}
The Grunwald-Letnikov based Fractional-order differ-integration\textit{G-L} defines the fractional integration and differentiation in a unified way:
\begin{equation}
\label{GL_DI}
  _{t_{in}}^{GL}d_t^{\alpha}f\left ( t \right )=\lim_{h\to 0 }h^{-\alpha}\mathlarger{\sum}_{j=0}^{{\left \lfloor \dfrac{t-t_{in}}{T_s}  \right \rfloor }}(-1)^j\binom{\alpha}{j}f(t-j\,\,h),
 \end{equation}
\noindent where $h$ is the sampling time, $ \mathlarger{\binom{\alpha}{j}}= \dfrac{\Gamma(\alpha+1)}{\Gamma(j+1)\Gamma(\alpha-j+1)}$ and $\left \lfloor . \right \rfloor$ describes the floor function and denotes the biggest integer smaller or equal to the argument.
\end{definition}
\begin{remark}
Suppose $f \in C^ {\left \lfloor \alpha \right \rfloor +1}$, the \textit{G-L} derivative is equivalent to the \textit{R-L} derivative (see \cite{diethelm2010analysis}).
\end{remark}
%%%%%%%%%%%%%%%%%%%%%%%%%%%%%%%%%%%%%%%%%%%%%%%%%%%%%%%%%%%%%%%%%%%%%%%%%%%%%%%%%%%%%%
\subsection{Aberration phenomenon}
This section focuses on the proposition of the so-called aberration phenomena that represent the effect of the pre-initial processes on a fractional-order system's response. To do so, we simulate a simple system that model the internal force of an axially loaded viscoelastic bar, \cite{efe2011fractional}. In the following illustration example, $u(t)$ is the input denoting the longitudinal force, and $y(t)$ is the output representing the elongation. They satisfy the following fractional-order relationship:
\begin{equation}
   _{t_0}D_t^{0.5} y(t)=u(t)
\end{equation}
Assume the force $u(t)$, the input of the system, is zero before the absolute zero instant of the system, and the initial instant is taken as $t_{in}=5$. We excite the system by three different inputs $U_0$, $U_1$, and $U_2$ that are specified as follows:
\begin{equation}
\begin{matrix}
U_0\left\{\begin{matrix}
0, \ \ t\leqslant 0\\
0, \ \ 0<t<5 \\
1, t\geqslant 5
\end{matrix}\right.,
 & U_1\left\{\begin{matrix}
0, \ \ t\leqslant 0\\
t, \ \ 0<t<5 \\
1, t\geqslant 5
\end{matrix}\right., 
& U_2\left\{\begin{matrix}
0, \ \ t\leqslant 0\\
\dfrac{1}{4}t^4, \ \ 0<t<5 \\
1, t\geqslant 5
\end{matrix}\right.
\end{matrix}
\end{equation}
\begin{figure*}[!t]
	\centering
	\includegraphics[width=1\textwidth]{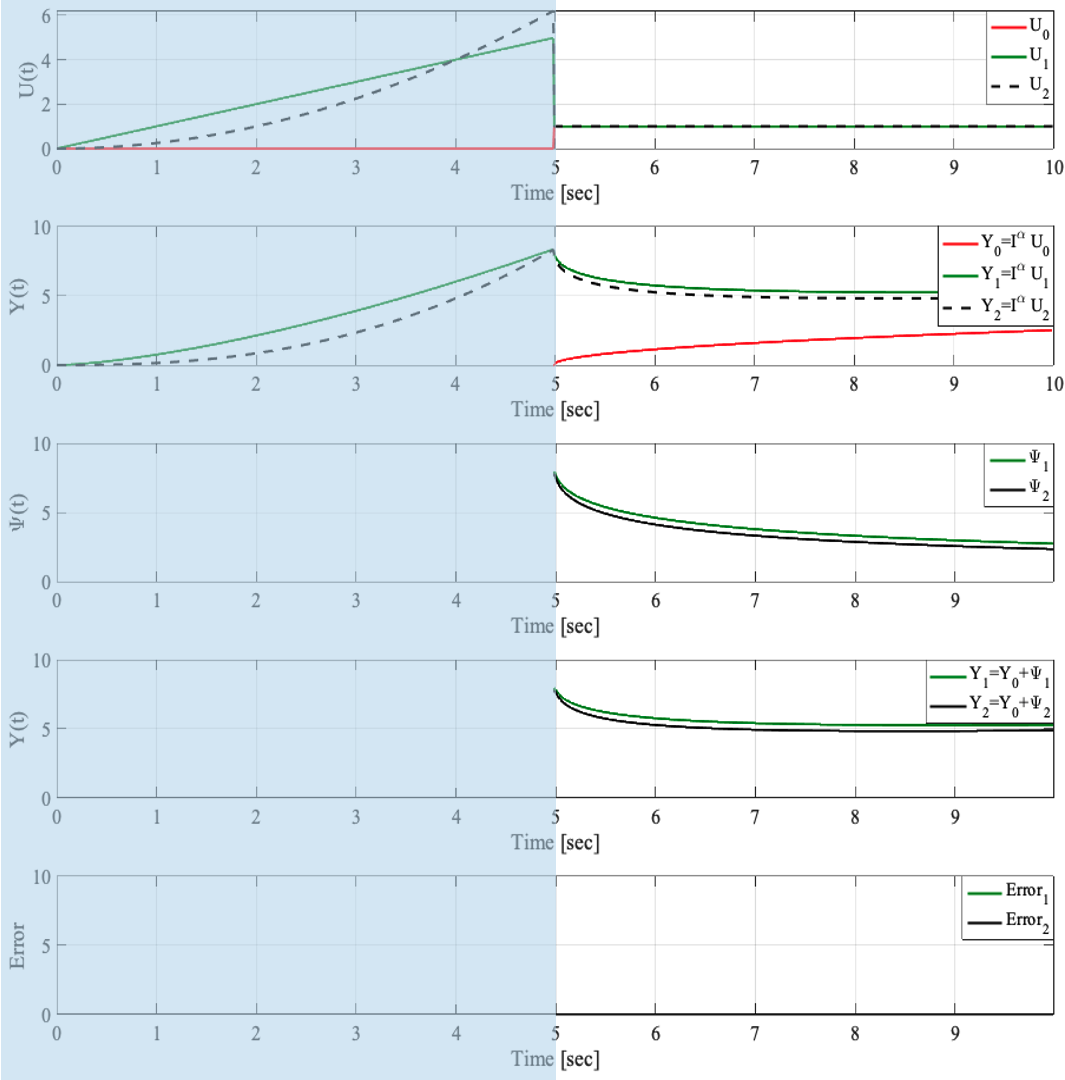}
	\caption{ The aberration phenomenon.}
	\label{figin3}
\end{figure*}
The corresponding outputs ($Y_0$, $Y_1$, and $Y_2$) along with the initialization function, as developed in the previous section, are shown in Figure ~\ref{figin3}. It is clear from this graph that although $Y_1$ and $Y_2$ have the same values at the initial time $t_{in}=5$ and they are excited with the same input afterward, they don't have the same response later. In addition, from the output $Y_0$ that corresponds to a non-initialized response as $u(t)$ is equal to zero over the pre-initialized process $t \in [0 \ \ 5]$. From all these outputs, we can perceive the effect of the pre-initial process. Accordingly, in the simulation of the fractional-order system, the ignorance of infinite states at $t_{in}$ or the improper initialization of the states before the initial instant might cause an aberration of the output. On the one hand, the noticed aberration phenomenon reflects the complexity and the difficulty of the initial value problem. On the other hand, it illustrates the great challenge of analyzing and controlling fractional-order systems. Commonly, to overcome the negative consequences of the aberration, former studies are used to transform the problems of the non-initial value into zero initial value ones.  Hence, it is essential to seek the inner essence of this aspect. Furthermore, it is primordial to provide adapted procedures to select the exact time-varying history function that leads to the correct response and accurate estimation of the system's states.
\subsection{Time-varying function based-initialization of FOS}
The main idea of the history function based-initialization is that the initialization time-varying function $\Psi(t)$ should bring out the past history. Otherwise stated, the history function entails the effect of fractionally integrating the function from its birth \cite{das2011initialized,belkhatir2018parameters}.\\
\noindent Assume that $f(t)$ was born at $t=t_{abs}$, that is $f(t)=0$ for all time less than equal to $t_{abs}$.
%. i.e. $f(t)=0$, $t<t_{abs}$.
Then the time period between $t_{abs}$ and $t_{in}$ may be considered as history if the fractional-order integration starts at $t=t_{in}$ and $f(t)=f_{in}(t)$ $\forall \,\,\, t\in[t_{abs}\,\,\, t_{in}]$. %$f_{in}(t)$ is called history function.%
The fundamental idea is that the fractional-order integral ($_{t_{in}}d_t^{-\alpha} f(t)$), should be properly initialized so that it should function as continuation of integral starting at $t = t_{abs}$. Hence an initialization function, $\Psi$ must be added to ($ _{t_{in}}d_t^{-\alpha} f(t)$), so that the fractional-order integral starting at $t = t_{in}$ should be identical to the result starting at $t=t_{abs}$ for $t \geq t_{in}$. The history function $\Psi$ has the effect of allowing the function $f(t)$ and its derivatives to start at a value other than $t_{in}$, namely $_{t_{abs}}D_{t_{in}}^{-\alpha}f(t)\mid_{t=t_{in}}$ , and continues to contribute to differ-integral response after $t = t_{in}$. That is, a function of time is added to the uninitialized integral, (not just a constant at $t=t_{in}$). The above argument can be formulated as follows: for $t>t_{in}$
\begin{equation}
\label{initialization}
  \Psi(f_{in},\alpha,t_{abs},t_{in},t)=_{t_{abs}}d_t^{-\alpha} f(t)- _{t_{in}}d_t^{-\alpha} f(t).
\end{equation}
\textcolor{black}{
Generally, we differentiate two types of initialization:
\begin{itemize}
\item  \textbf{Side-initialization:} Fully arbitrary initialization may be applied to the differ-integral operator at time $t = t_{in}$ .
\item \textbf{Terminal initialization:} It is assumed that the differ-integral operator can be initialized (charged) by effectively differ-integrating prior to the start time, $t = t_{in}$. 
\end{itemize}
Figure ~\ref{figin1} demonstrates the concept of initialization as a block diagram of a signal flow graph.
\begin{figure*}[!t]
	\centering
	\includegraphics[width=.85\textwidth]{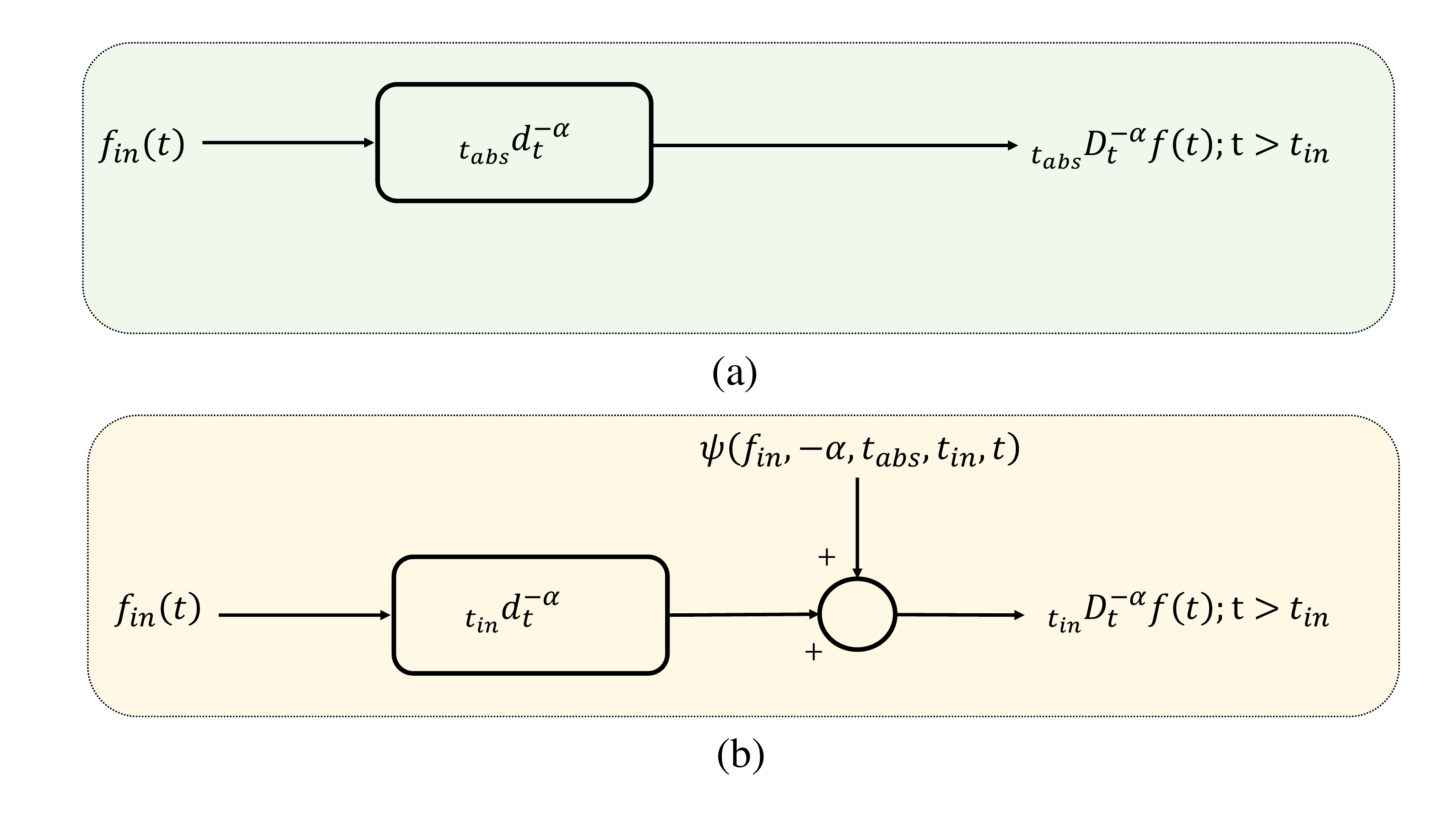}
	\caption{\textcolor{black}{ Signal flow graph for demonstrating initialization of fractional integration \cite{das2011initialized}. (a) illustrates the Side-initialization and (b) Terminal initialization}}
	\label{figin1}
\end{figure*}
%%%%%%%%%%%%%%%%%%%%%%%%%%%%%%%%%%%%%%%%%%%%%%%%%%%%%%%%%%%%%%%%%%%%%%%%%%%%%%%%
\subsubsection{History-function based-initialization of fractional-order integration}
\noindent Assume, the fractional-order integration of a function $f(t)$ starts a $t=c$ and takes place for all $t > t_{in} \geqslant t_{abs}$. The initialization process takes place during the period $t \in [t_{abs} \ \ t_{in}]$.
Generally the \textit{Riemann-Liouville (R-L)}  definition is admitted in the case where the differintegrand $f (t) = 0$ for all $t \leqslant t_{abs}$. The fractional-order integration based \textit{R-L} definition of $f(t)$ is: 
\begin{equation}
    _{t_{abs}}D_t^{-\alpha} f(t)=_{t_{abs}}d_t^{-\alpha} f(t) = \dfrac{1}{\Gamma(\alpha)}\int_{t_{abs}}^t(t-\tau)^{\alpha-1}f(\tau)d\tau, 
\end{equation}
where ($\alpha \geqslant0$ and $t>t_{abs}$ subject to $f(t)=0$ for all $t\leqslant t_{abs}$. $\Psi( f ,-\alpha, t_{abs},t_{abs},t) = 0$\\
\noindent The following definition of fractional-order integration will apply generally (at
any $t > c$ ):
\begin{equation}
    _{c}D_t^{-\alpha} f(t)=\dfrac{1}{\Gamma(\alpha)}\int_c^t(t-\tau)^{\alpha-1}f(\tau)d\tau +\Psi(f,-\alpha,a,c,t) 
\end{equation}
where ($\alpha \geqslant0$, $t>t_{abs}$, $t_{in}>t_{abs}$ and $f(t)=0$ for all $t\leqslant t_{abs}$.\\
\noindent The function $\Psi(f,-\alpha,t_{abs},t_{in},t)$ is called the initialization function and will be chosen such that
\begin{equation}
    _{t_{abs}}D_t^{-\alpha}f(t)=_{t_{in}}D_t^{-\alpha}f(t), \ \ t>t_{in}
\end{equation}
The above condition gives:
\begin{equation}
    \dfrac{1}{\Gamma(\alpha)}\int_{t_{abs}}^t(t-\tau)^{\alpha-1}f(\tau)d\tau =\dfrac{1}{\Gamma(\alpha)}\int_{t_{in}}^t(t-\tau)^{\alpha-1}f(\tau)d\tau +\Psi(f,-\alpha,t_{abs},t_{in},t) 
\end{equation}
Hence 
\begin{equation}
\begin{matrix}
     \Psi(f,-\alpha,t_{abs},t_{in},t)= \dfrac{1}{\Gamma(\alpha)}\int_{t_{abs}}^t\left(t-\tau)^{\alpha-1}f(\tau\right)d\tau -\dfrac{1}{\Gamma(\alpha)}\int_{t_{in}}^t(t-\tau)^{\alpha-1}f(\tau)d\tau
    \end{matrix}
\end{equation}
Since $\int_{t_{abs}}^tg(\tau)d\tau=\int_{t_{abs}}^{t_{in}}g(\tau)d\tau+\int_{t_{in}}^t g(\tau)d\tau$ we get:
\begin{equation}
     \Psi(f,-\alpha,t_{abs},t_{in},t)=_{t_{abs}}D_{t_{in}}^{-\alpha}=\dfrac{1}{\Gamma(\alpha)}\int_{t_{abs}}^{t_{in}}\left(t-\tau)^{\alpha-1}f(\tau\right)d\tau 
\end{equation}
This expression for $\Psi(t)$ gives 'terminal initialization', and also brings out in the definition of fractional integral the effect of the past 'history', namely the effect of fractionally integrating the $f(t)$ from $t_{abs}$ to $t_{in}$. This effect is also called terminal charging.
%%%%%%%%%%%%%%%%%%%%%%%%%%%%%%%%%%%%%%%%%%%%%%%%%%%%%%%%%%%%%%%%%%%%%%%%%%%%%%%%%%%%%%%%
\subsubsection{History-function based-initialization of fractional-order derivative}
Unlike the integer-order derivative that is considered a local quantity, the fractional-order derivative (\textit{FD}) is a non-local operator and has a history. Furthermore, as \textit{FD} contains fractional-order integration, its evaluation requires the initialization process. 
\noindent By considering the integer-order derivative as a particular case of the fractional-order derivative, a generalization of the integer-order derivative concept also calls for initialization. Accordingly, a generalized integer-order differentiation with initialization can be defined as:
\begin{equation}
     _{t_{in}}D_t^{m} f(t)=\dfrac{d^m}{dt^m}f(t)+\Psi(f,m,t_{abs},t_{in},t), t>t_{in}
\end{equation}
where $m$ is an integer corresponding to the differentiation order and $\Psi(f,m,t_{abs},t_{in},t)$ is the initialization function.
\noindent Assume a fractional-order differentiation; $\alpha \geqslant 0$.  A function $f(t)$ is born at $t = t_{abs}$ and before that the value is zero ($f(t)=0$ for $t\geqslant t_{abs}$). $\alpha=m-\beta$ equivalently $m$ is the integer just greater than the fractional order $\alpha$, by amount $\beta$.   The differentiation starts at $t > t_{in}$.\\ 
\noindent A non-initialized fractional-order derivative can be expressed as:
\begin{equation}
    _{t_{abs}}d_t^{\alpha}f(t)=_{t_{abs}}D_t^{m} \ _{t_{abs}}D_t^{-\beta}f(t)
\end{equation}
\noindent Similar to the fractional-order integration in this case the initialization function $\Psi( f ,-\beta, t_{abs},t_{abs},t) = 0$.\\
\noindent If we consider the $h(t)= _{t_{abs}}D_t^{-\beta}f(t)$ i.e. fractional integral of function starting at $t_{abs}$ with initialized term $\Psi(h,m,t_{abs}, t_{abs},t) = 0$; the initialized fractional derivative looks and defined as for $\alpha \geqslant 0$, $t >t_{in}\geqslant t_{abs}$.
\begin{equation}
    _{t_{in}}D_t^{\alpha}f(t)=_{t_{in}}D_t^{m} f(t) \ _{t_{in}}D_t^{-\beta}f(t)
\end{equation}
Accordingly, the history-function based-initialization of the fractional-order derivative is generally similar to that obtained with initialization of the fractional-order integral. In fact, it requires the following as the fractional-order integral: 
\begin{figure*}[!t]
	\centering
	\includegraphics[width=1\textwidth]{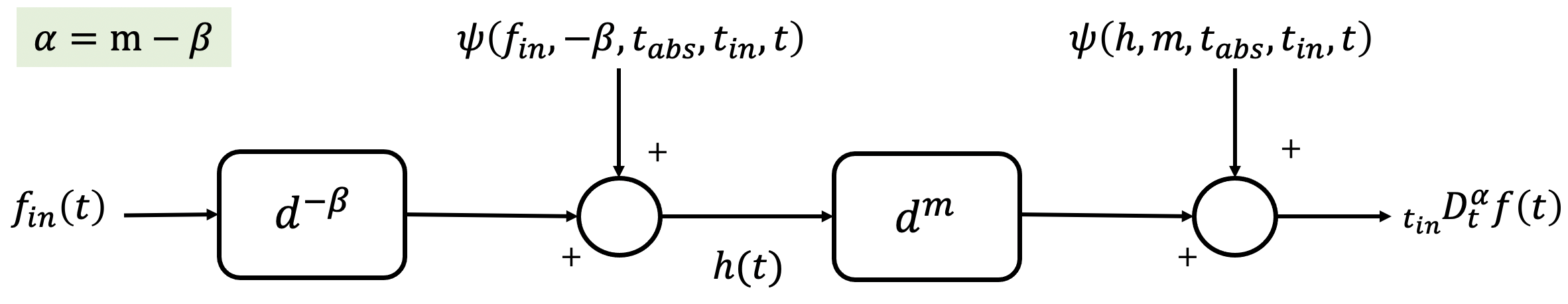}
	\caption{ Initialization of fractional derivative.}
	\label{figin2}
\end{figure*}
\begin{equation}
    _{t_{abs}}D_t^{\alpha}f(t)=_{t_{in}}D_t^{\alpha}f(t)
\end{equation}
%%%%%%%%%%%%%%%%%%%%%%%%%%%%%%%%%%%%%%%%%%%%%%%%%%%%%%%%%%%%%%%%%%%%%%%%%%%%%%%%
For all $t>t_{in}\geqslant t_{abs}$. Specifically this requires compatibility of the derivatives starting at $t = t_{abs}$ and $t = t_{in}$ for all $t > t_{in}$.\\
\noindent Therefore it follows that:
\begin{equation}
   _{t_{in}}D_t^{m} \ _{t_{in}}D_{t}^{-\beta} f(t)=_{t_{abs}}D_t^{m} \ _{t_{abs}}D_{t}^{-\beta} f(t)
\end{equation}
\noindent Expanding the fractional-order integrals with the initialization function introduced in the previous section we obtain:
\begin{equation}
\begin{matrix}
   _{t_{in}}D_t^{m}\left(\dfrac{1}{\Gamma(\beta)}\int_{t_{in}}^t(t-\tau)^{\beta-1}f(\tau)d\tau+\Psi(f,-\beta,t_{abs},t_{in},t)\right)= \\ _{t_{abs}}D_t^{m} \left(\dfrac{1}{\Gamma(\beta)}\int_{t_{abs}}^t(t-\tau)^{\beta-1}f(\tau)d\tau+\Psi(f,-\beta,t_{abs},t_{abs},t)\right)
   \end{matrix}
\end{equation}
For, $t>t_{in}$ and $\Psi(f,-\beta,t_{abs},t_{abs},t)=0$. Using the definition of generalized integer order derivative as defined above we get:
\begin{equation}
\begin{matrix}
   \dfrac{d^m}{dt^m}\left\{\dfrac{1}{\Gamma(\beta)}\int_{t_{in}}^t(t-\tau)^{\beta-1}f(\tau)d\tau+\Psi(f,-\beta,t_{abs},t_{in},t)\right\}+\Psi(h_1,m,t_{abs},t_{in},t)= \\ \dfrac{d^m}{dt^m} \dfrac{1}{\Gamma(\beta)}\int_{t_{abs}}^t(t-\tau)^{\beta-1}f(\tau)d\tau+\Psi(h_2,m,t_{abs},t_{abs},t)
   \end{matrix}
\end{equation}
where $h_1=_{t_{in}}D_t^{-\beta}f(t)$ and  $h_2=_{t_{abs}}D_t^{-\beta}f(t)$. 
\noindent The integer order derivative is initialized at $t = a$ thus $\Psi(h_2,m,t_{abs},t_{abs},t)=0$. After rearranging the integrals we get: 
\begin{equation}
    \Psi(h_1,m,t_{abs},t_{in},t)=\dfrac{d^m}{dt^m}\left(\dfrac{1}{\Gamma(\beta)}\int_{t_{abs}}^{t_{in}}(t-\tau)^{\beta-1}f(\tau)d\tau-\Psi(f,-\beta,t_{abs},t_{in},t)\right)
\end{equation}
The above expression is the requirement for the initialization of the derivative in general.
\noindent Under the condition of terminal charging of the fractional-order integral, the initialization function of fractional integration as defined and derived earlier is:
\begin{equation}
\Psi(f,-\beta,t_{abs},t_{in},t)=\dfrac{1}{\Gamma(\beta)}\int_{t_{abs}}^{t_{in}}(t-\tau)^{\beta-1}f(\tau)d\tau
\end{equation}
Hence the  $\Psi(h_1,m,t_{abs},t_{in},t) = 0 $.
Figure ~\ref{figin2} illustrates the initialization concept for fractional derivative.}
%%%%%%%%%%%%%%%%%%%%%%%%%%%%%%%%%%%%%%%%%%%%%%%%%%%%%%%%%%%%%%%%%%%%%%%%%%%%%%%%
\subsubsection{Grunwald-Letnikov based fractional-order differ-integrals initialization concept}
\begin{definition}
\label{def_GL_psi}
The formulations of the time-varying history function using the most known fractional-order differ-integration definitions namely \textit{RL}, \textit{C} and \textit{GL} can be observed in \cite{lorenzo2007initialization}. Here we present only the \textit{G-L} based-one as our proposed method is based on it. 
\begin{equation}
\label{initialization_GL}
\begin{matrix}
\Psi_{GL}(f_{in},\alpha,t_{in},t_{abs},t)= \underset{{h \rightarrow 0}}{\lim} \left \{\dfrac{h^{-\alpha}}{\Gamma(-\alpha)} \mathlarger{\sum}_{j=0}^{N_3-1}\dfrac{\Gamma\left(N_1-1-\alpha-j\right)}{\Gamma\left(N_1-j\right)}f[t-(N_1-1-j)h]\!\!\right \},
\end{matrix}
\end{equation}
\noindent where $ {h={(t-t_{abs})}\setminus{N_1}}$ is the time step, $ t>t_{in}>t_{abs}$ and $N_1$ and $N_3$ are integers such that:
\begin{equation}
    \label{ratio}
    {N_3=\left( \dfrac{t_{in}-t_{abs}}{t-t_{abs}}\right)N_1}
\end{equation}
\end{definition}
\section{Problem Formulation}
Let us consider the following FOS:
\begin{equation}
    \label{FOS1}
    y(t)+\mathlarger{\sum}^N_{i=1}a_i\,\,\,\, _{t_{in}}D^{\alpha_i}_ty(t)=b \,\, u(t), \,\, \,\, \,\,\,\,\,\,\,\,\,\,t \in [t_{in} \,\,\,\,\,\,T]
\end{equation}
where $ y(t): [t_{in} \,\,\,T] \rightarrow \mathbb{R}$ is the output, $ u(t): \,\,\, [t_{in} \,\,\,T] \rightarrow \mathbb{R}$ is the input. $(a_1\ a_2\ \dots a_N)$ and $b$ are real parameters. $\alpha_i \in \wp=(n_{i-1},n_i)$ \textcolor{black}{\footnote{\textcolor{black}{$\wp = \prod^{N}_{i=1} \wp_i=\{(a_1,a_2\dots , a_N)| \forall i=1,2,\dots ,N, a_i \in \wp_i\}$ is the generalized Cartesian product of $N$ sets $\wp_1, \wp_2 \dots \wp_N $. }}}, with $n_i \in \mathbb{N}^*$ and $i=1,2, \dots, N$ are the fractional differentiation orders. They are assumed to be as follows: ${0\leq\!\!\alpha_1\leq\!\!\alpha_2\leq\dots\leq\!\!\alpha_n}$, i.e., $  {n_i<n_{i+1}}$ for ${i=1,2,\dots ,N-1}$. We substitute the initialized fractional derivative \eqref{pre2} in \eqref{FOS1} which leads the following equation:
\begin{equation}
    \label{FOS2}
    y(t)+\mathlarger{\sum}^N_{i=1} \textcolor{black}{a_i\,\,\,\,}  \left[ _{t_{in}}d^{\alpha_i}_t y(t)+ \Psi_i(f_{in},\alpha_i,t_{abs},t_{in},t)\right]\!\!=b \,\, u(t),
\end{equation}
$ f_{in}(t): \,\,\,[t_{abs} \,\,\,t_{in}] \rightarrow \mathbb{R}$ corresponds to the history function of $y(t)$ based-initialization.  

\noindent ${\Psi=(\,\,\,\Psi_1(f_{in},\alpha_1,t_{abs},t_{in},t) \,\,\,\,\,\, \Psi_2(f_{in},\alpha_2,t_{abs},t_{in},t)\,\,\, \dots}$ ${\,\,\,\Psi_N(f_{in},\alpha_N,t_{abs},t_{in},t)\,\,\,)^{tr}}$ denotes the vector of initialization functions and and are defined as in (\ref{initialization}).

\noindent We denote the vectors $p$ and $\alpha$ as: $p=(a_1\ a_2\ \dots a_N, b)^{tr}$, $\alpha=(\alpha_1\ \alpha_2\ \dots \alpha_N)^{tr}$ and their estimates as  $\hat p=(\hat a_1\ \hat a_2\ \dots \hat a_N, \hat b)^{tr}$ and $\hat{\alpha}=(\hat{\alpha}_1\ \hat{\alpha}_2\ \dots \hat{\alpha}_N)^{tr}$  respectively.
\noindent $(\cdot)^{tr}$ denotes the transpose of the row vector.

\noindent In this work, we are interested in solving the following estimation problem (EP) that accounts for the initialized fractional-order derivative:\\

\noindent (EP) $\left\{ 
\begin{tabular}{l} 
\textit{Given the output signal $y(t)$ and $u(t)$, $t \in [t_{in},T]$,}
\textit{design a history function based-initialization,}\\
\textit{$f_{in}(t)$, $t \!\!\in [t_{abs},t_{in}]$ jointly with the simultaneous}
\textit{estimation of the unknown  parameters and}\\
\textit{ the fractional orders, denoted $(\hat{p}, \hat{\alpha})$}
\end{tabular}\right.$
%%%%%%%%%%%%%%%%%%%%%%%%%%%%%%%%%%%%%%%%%%%%%%%%%%%%%%%%%%%%%%%%%%%%%%%%%%%%%%%%
\section{Reformulation of the Estimation Problem in Discrete Space}
\textcolor{black}{
In this section, we reformulate the initialized estimation problem in the discrete space using the Grunwald-Letnikov derivative definition given in Definition \ref{def_GL} and the history function based $GL$ given in Definition \ref{def_GL_psi}. In the following, we present the different steps of the problem formulations:\\
\noindent For a given small enough sampling time $h$ such that ${h={(t-t_{abs})}\setminus{N_1}}$,\\
\noindent ${N_3=\left( {t_{in}-t_{abs}}\setminus{t-t_{abs}}\right)N_1}$ and $K=N_1-N_3$. In the discrete space the system (\ref{FOS2}) can be written as follows $\forall \,\,\, k=1,2,... K$
 \begin{equation}
    \label{FOS2_disc}
    y(t_k)+\mathlarger{\sum}^N_{i=1}a_i\,\,\,\,  \left[ _{t_{in}}d^{\alpha_i}_t y(t_k)+ \Psi_i(f_{in},\alpha_i,t_{abs},t_{in},t_k)\right]\!\!=b \,\, u(t_k),
\end{equation}
where
\begin{equation}
\label{psi}
\begin{matrix}
\Psi_i(f_{in},\alpha_i,t_{in},t_{abs},t_k)= \dfrac{1}{h^{\alpha_i}} \mathlarger{\sum}_{n=0}^{N_3-1}C_n^{\alpha_i}f_{in}[t_k-(N_1-1-n)h].
\end{matrix}
\end{equation}
where $C_j^{\alpha_i} (j = 0, 1, \dots N_1)$ are the binomial coefficients recursively computed using the following formula:
\begin{equation}\label{bionom}
C_0^{\alpha_i} =1, \,\,\,\,\,\,\,  C_j^{\alpha_i}=\left(1-\dfrac{1+\alpha}{j}\right)C_{j-1}^{\alpha_i}.
\end{equation}
Substituting (\ref{psi}) in (\ref{FOS2_disc}) we obtain: 
 \begin{equation}
    \label{FOS2_disc_1}
    y(t_k)+\mathlarger{\sum}^N_{i=1} a_i\,\,\,\, \left[  _{t_{in}}d^{\alpha_i}_t y(t_k)+ \dfrac{1}{h^{\alpha_i}}\mathlarger{\sum}_{n=0}^{N_3-1}C_n^{\alpha_i}f_{in}[t_k-(N_1-1-n)h]\right]\!\!=b \,\, u(t_k).
\end{equation}
Based on Definition \ref{def_GL} and substituting $d^{\alpha_i}_t y(t_k)$ by its formula described in (\ref{GL_DI}) we obtain the following:
 \begin{equation}
    \label{FOS2_disc_2}
    y(t_k)+\mathlarger{\sum}^N_{i=1}  a_i\,\,\,\, \left[ \dfrac{1}{h^{\alpha_i}}\,\displaystyle{\sum_{j=0}^{N_1}} C_j^{\alpha_i} y(t_{k-j})+ \dfrac{1}{h^{\alpha_i}}\mathlarger{\sum}_{n=0}^{N_3-1}C_n^{\alpha_i}f_{in}[t_k-(N_1-1-n)h]\right]\!\!=b \,\, u(t_k).
\end{equation}
By rearranging (\ref{FOS2_disc_2}) we obtain:
 \begin{equation}
    \label{FOS2_disc_3}
    \mathlarger{\sum}^N_{i=1}  a_i \,\,\, A_i(t_k) + b \,\, u(t_k)= y(t_k)
\end{equation}
where 
\begin{equation}
    \label{Ai}
   A_i(t_k)=-\dfrac{1}{h^{\alpha_i}}\left[\displaystyle{\sum_{j=0}^{N_1}} C_j^{\alpha_i} y(t_{k-j})+ \mathlarger{\sum}_{n=0}^{N_3-1}C_n^{\alpha_i}f_{in}[t_k-(N_1-1-n)h]\right]
\end{equation}
Accordingly, the reformulated discrete initialized discretized estimation problem (IDEP) is given in the following proposition.}
\begin{proposition}
For a given small enough sampling time $h$, the estimation problem (EP) can be reformulated in the discrete space as follows: 
\begin{equation}\label{estimate}
    F( \alpha;N_3) \cdot \,\,\, p = R,  
    %\hspace{0.2cm} \forall t_k \geq t_{in}
\end{equation}
where $p$ is the vector of unknown parameters and $R\in \mathbb{R}^{K \times 1}$ such that $K=N_1-N_3$, the number of samples within $[t_{in}, T]$, $R_k=y(t_k), \,\,\, k=1,2,\dots, K$. \\
The function $F( \alpha;N_3) \in \mathbb{R}^{K \times (N+1)}$ that is a non-linear function with respect to the fractional differentiation orders and the length of the initialization memory is given as follows:
for $\forall \,\,\,\, k = 1: K$
\begin{align}
\label{eq8}
F_{ki}(\alpha_i;N_3) = \left\{ 
\begin{array}{lll} 
A_i(t_k)=-\dfrac{1}{h^{\alpha_i}}\left[\displaystyle{\sum_{j=0}^{N_1}} C_j^{\alpha_i} y(t_{k-j})+ \mathlarger{\sum}_{n=0}^{N_3-1}C_n^{\alpha_i}f_{in}[t_k-(N_1-1-n)h]\right],  \\ \forall   i=1:N \\
u(t_k),  \hspace{.2cm} for  \,\, i=N+1, 
\end{array}
\right.
\end{align}
\end{proposition}
\begin{remark}
It is worth noting that the reformulated problem (\ref{estimate}) is expressed as a linear combination of nonlinear functions. The unknown variables in the objective function based on in (\ref{estimate}), can be separated into two disjoint sets: linear variables, $p$, and nonlinear variables, $(\alpha; N_3)$.This class of problems is known as the separable nonlinear least squares problem \cite{borges2009full}.
\end{remark}
\section{Estimation algorithm}
\begin{figure}[!t]
        \centering
        \includegraphics[width=1\linewidth]{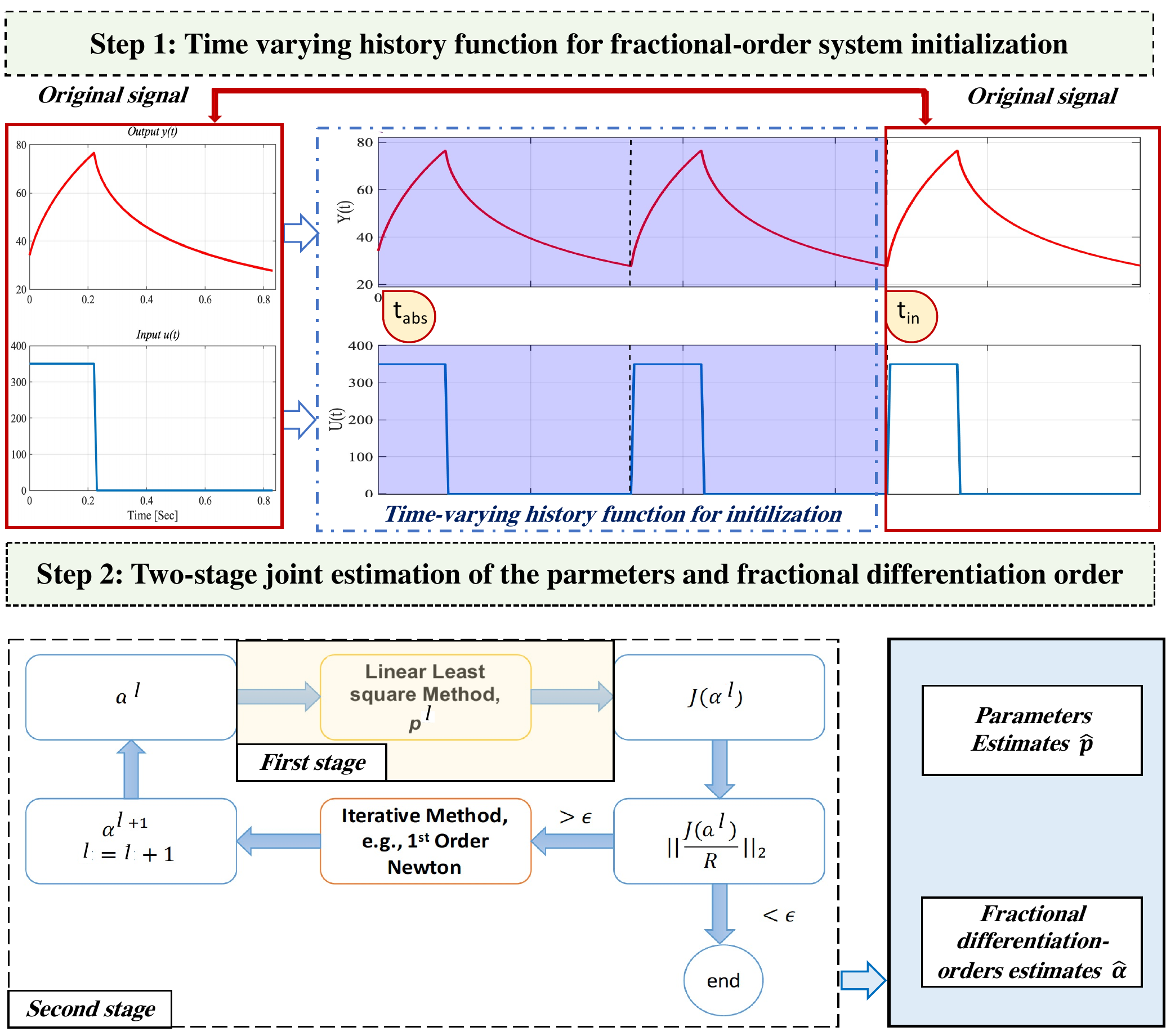}
\caption{A flowchart showing the two main steps of initialized estimation algorithm for parameters and fractional differentiation orders of fractional-order systems.}
\label{fig1}
\end{figure}
This section introduces the main two steps of the proposed method to solve (EP). The first step designs a time-varying history function to initialize the output signal of the FOS (\ref{FOS1}). The second step is an iterative hybrid approach to solve the estimation problem of the unknown parameters and fractional differentiation orders, (\ref{estimate}). This step contains two stages. The first stage solves the parameters' estimation problem at each iteration using a linear least square method by solving a linear system of equations; the second stage solves a nonlinear system of equations using Newton's method to find estimates for the fractional differentiation orders.
The main steps of the proposed method which are explained in the following are depicted in Figure \ref{fig1}.
\subsection{Step 1: Design of the time-varying history function}
The key hypothesis in step 1 is to propose an initialization function, $f_{in}(t)$, in the estimation process of the unknown parameters and fractional differentiation orders. The choice of $f_{in}(t)$ depends on the available measured output $y(t)$ where duplicates of $y(t)$ is considered. This choice reduces the infinite-dimensional space of possible initialization functions into a finite parametric space where the remaining degree of freedom is the number of output cycles to be duplicated. A schematic picture of the designed initialization framework (function) is depicted in Figure \ref{fig1}, Step 1.
\subsection{Step 2: Two-stage Estimation Algorithm for Parameters and Fractional Differentiation Order}
Let the fractional derivatives of the output signal, $y(t)$, be initialized with a time-varying function $f_{in}(t)$ as designed in the previous step.
In this study, we suppose that $N_3$ is known. Accordingly, in this situation the set of parameters $p$ enters linearly while the set of parameters $\alpha$ non-linearly in the following predictor: 
\begin{equation}
\label{y_hat}
    \hat y(t\mid p,\alpha)= p^{tr} F^{tr}(t,\alpha;N_3).
\end{equation}
The identification criterion then becomes
\begin{equation}
\label{criterion}
    V_{K}(p,\alpha)=\mathlarger{\sum}^K_{k=1}\mid y(t_k)-p^{tr}F^{tr}(t_k,\alpha;N_3)\mid^2=\mid R-F(\alpha;N_3) p \mid,
\end{equation}
where we introduced matrix notation, analogously to (\ref{estimate}). Then, for a given $\alpha$, this criterion is the least square criterion and minimized w.r.t. $p$ by
\begin{equation}
\label{estimate_y}
    \hat p=\left[F^{tr}(\alpha;N_3)F(\alpha;N_3)\right]^{-1} F^{tr}(\alpha;N_3) \,\,\,R
\end{equation}
By inserting (\ref{estimate_y}) into the identification criterion (\ref{criterion}) the non-linear problem with repst to $\alpha$ can be defined as:
\begin{equation}
    \underset{\alpha}{\min} \mid R-P(\alpha)R \mid^2=\underset{\alpha}{\min} \left | \left[I_{K \times K}-P(\alpha)\right] R \right |^2 \coloneqq \underset{\alpha}{\min}  \mid J(\alpha;N_3) \mid^2
\end{equation}
where
\begin{equation}
    \label{eqJ}
    J(\alpha;N_3)= \left[I_{K \times K}-P(\alpha)\right]R
\end{equation}
with $I_{K \times K}$ being the identity matrix of dimension $K$ and $P(\alpha)$ is defined as:
\begin{equation}
        P(\alpha)= F(\alpha;N_3)\left[F^{tr}(\alpha;N_3)F(\alpha;N_3)\right]^{-1} F^{tr}(\alpha;N_3)
\label{eqP}
\end{equation}
%%%%%%%%%%%%%%%%%%
\begin{remark}
The estimate $\hat p$ is obtained by inserting the minimizing $\alpha$ into (\ref{eqJ}). The matrix $P$ is a projection matrix: $P^2=P$. The described method is called separable least squares since the least-square part has been separated out, and the problem is reduced to a minimization problem of lower dimension.
\end{remark}
%%%%%%%%%%%%%%%%%
\noindent An estimate $\hat \alpha^l$ of the vector of fractional differentiation orders are computed iteratively by solving the nonlinear problem:
\begin{equation}
    \label{sol3}
    J (\alpha;N_3) =0
\end{equation}
with $J \in \mathbb{R}^{K \times 1}$, which is defined in Eqt. (\ref{eqJ}), using the following Newton update law:
\begin{equation}
\label{newt}
    \hat{\alpha}^{l+1}=\hat{\alpha}^l-[J'(\hat{\alpha}^l;N_3)]^{-1} J(\hat{\alpha}^l;N_3),
\end{equation}
 where $J'(\alpha;N_3) \in \mathbb{R}^{K \times N}$ is the Jacobian matrix of the nonlinear function $J$.
%%%%%%%%%%%%%%%%%%%%%%%%%%%%%%%%%%%%%%%%%%%%%%%%%%%%%%%%%%%%%%%%%%%%%
\begin{proof}
The system of equations (\ref{estimate}) is solved in two stages. While in the first step we use the linear least square method to estimate the parameters, in the second step we use \textit{Newton} iterative method to estimate the fractional differentiation orders.\\
 \textbf{First stage:} From (\ref{estimate}) we have
\begin{equation}
\label{sol1}
        F( {\alpha^l};N_3) \cdot \,\,\, p = R,
\end{equation}
For any given $\alpha$, we denote $\hat{p}$ the solution of (\ref{sol1}) for the unknown vector of parameters $p$. Thus, for $\alpha= \hat{\alpha}^l$ the system of equations(\ref{sol1}) can be written as in (\ref{estimate_y}).\\
 \textbf{Second stage:} From (\ref{estimate}) and the first stage we have
 \begin{equation}
 \label{sol2}
             F( {\alpha};N_3) \cdot \,\,\, \hat p = R,
 \end{equation}
 where $F( {\alpha};N_3)$ and $R$ are introduced in (\ref{eq8}). We propose to use the iterative first-order Newton method to solve(\ref{sol2}), with a solution at iteration $l$ denoted $\hat\alpha^l$. The Newton update law is given in (\ref{newt}). At each iteration, (\ref{sol3}) follow directly from (\ref{sol2}).
 
\noindent The Jacobian matrix ($J'(\alpha)$) needs to be evaluated at each iteration $l$. Thus, we first substitute $\alpha$ by $\alpha^l$, and then we use it in the updated law (\ref{newt}).
\end{proof}
\noindent The main steps of the proposed method are given in Algorithm 1. 
\begin{algorithm}[!h]
\caption{ Proposed two-stage estimation method}
\begin{algorithmic}[1]
\State initialize $l= 0$, give an initial guess to $\alpha^l$,
\State compute $\hat p$ using equation (\ref{estimate_y}) ,
\State compute $ J(\alpha^l )$ using equations (\ref{eqJ}), 
\State \textbf{if} $ \parallel\dfrac{J(\alpha^l)}{R} \parallel\prec\epsilon $ then\footnote{$\epsilon$ is a user-defined threshold that takes small values.}
\State stop 
\State \textbf{else} 
\State update $\alpha^l$ using equations,
\State update the number of iterations $ l=l+1$, 
\State return to step 2
\State \textbf{end if}
\end{algorithmic}
\end{algorithm}
\vspace{-1cm}
%%%%%%%%%%%%%%%%%%%%%%%%%%%%%%%%%%%%%%%%%%%%%%%%%%%%%%%%%%%%%%%%%%%%%%%%%%%%%%%%%%%%%%%%%%%%%%%%%
\section{Convergence of the algorithm}
\subsection{Characterization of the Jacobian matrix}
\noindent In this part, we show the analytical derivation of the Jacobian matrix $J'(\alpha)$. The following lemmas are needed in the derivation of the Jacobian matrix and the convergence of the proposed algorithm.\\
\begin{lemma}
\label{lem1}
\cite{kurmayya2008moore}, Let $M \in R^{K\times N}$. $M^{tr}M$ is called the Gram matrix of $M$. The \textit{Moore-Penrose} inverse of $M$ is $M^+$ and we have:
\begin{equation}
    M^+=(M^{tr}M)^+M^{tr}=M^{tr}(M\,\,M^{tr})^+.
\end{equation}
\end{lemma}
\begin{lemma}
Let $M(\alpha)$ be a defined on an interval $I$ whose values are nonzero $n \times n$ matrices,  differentiable at $\alpha^*$.  Then the function $M^+$ is differentiable at $\alpha^*$ if an only if $rank(M(\alpha))$ is constant in some interval $|\alpha -\alpha^*|<\delta$. The derivative of the $ \dfrac{\partial M^+}{\partial\alpha} (\alpha^*)$ is given by:
\begin{equation}
\begin{array}{ll}
 \dfrac{\partial M^+}{\partial\alpha}=
    -M^+\left(\dfrac{\partial M}{\partial\alpha}\right)M^++M^+(M^+)^{tr}\left(\dfrac{\partial M^{tr}}{\partial\alpha}\right) e +
       e\left(\dfrac{\partial M^{tr}}{\partial\alpha}\right)(M^+)^{tr}M^+,
\end{array}
\end{equation}
where
$e= \left(I-M M^+\right)$ and $M, M^{tr}, M^+, \dfrac{\partial M}{\partial \alpha}$ stand for $M(\alpha^*)$, $M^{tr}(\alpha^*)$, $M^+(\alpha^*)$, $\dfrac{\partial M}{\partial \alpha}(\alpha^*)$ respectively.
\label{lem2}
\end{lemma}
\noindent Based on the formulation of $J(\alpha;N_3)$ in (\ref{eqJ}) and (\ref{eqP}) the elements of the Jacobian matrix $J'(\alpha;N_3)$ can be computed as:
$\forall \,\,\,$ $i=1:K$ and, $j=1:N$:
\begin{equation}
    J'_{ij}= \dfrac{\partial J_i}{\partial \alpha_j}=\mathlarger{\sum}_{l=1}^K \dfrac{\partial}{\partial \alpha_j}P_{il}(\alpha) R_l.
    \label{J'}
\end{equation}
Such that:
\begin{equation}
\label{P}
    P(\alpha)= F(\alpha)\,\,\, \overbrace{\left[\underbrace{F^{tr}(\alpha)F(\alpha)}_{Gram \,\,\,\,matrix}\right]^{-1} F^{tr}(\alpha)}^{F^+}.
\end{equation}
The matrix $F(\alpha)$ can be expressed as:
\begin{equation}
 F(\alpha)=  \begin{pmatrix}
&F_{11}(\alpha) & F_{12} (\alpha) & F_{13}(\alpha)  & \dots & F_{1(N+1)}(\alpha)&\\ 
&F_{21}(\alpha) & F_{22} (\alpha) & F_{23} (\alpha) & \dots & F_{2(N+1)}(\alpha)&\\ 
&F_{31}(\alpha) & F_{32} (\alpha) & F_{33} (\alpha) & \dots & F_{3(N+1)}(\alpha)&\\ 
&. & . & . &  \dots &  .&\\ 
& .& . & . & \dots    & . &\\ 
&. &.  &  .& \dots   &  .&\\ 
& F_{K1}(\alpha) & F_{K2} (\alpha) & F_{K3} (\alpha) & \dots & F_{K(N+1)}(\alpha)&
\end{pmatrix}
\end{equation}
%%%%%%%%%%%%%%%%%%%%%%% &
\noindent Based on \textbf{Lemma. \ref{lem1}}, we note:
\begin{equation}
    \label{B}
    B=F^+=\left[F^{tr}(\alpha)F(\alpha)\right]^{-1} F^{tr}(\alpha)=\left[F^{tr}(\alpha)F(\alpha)\right]^{+} F^{tr}(\alpha),
\end{equation}
is the \textit{Moore-Penrose} inverse matrix. Accordingly, (\ref{P}) can be written as:
\begin{equation}
    P(\alpha)= F(\alpha)\,\,\,  F^+(\alpha)= F(\alpha)\,\,\, B (\alpha),
\end{equation}
where $B(\alpha)$ is defined as:
\begin{equation}
    B(\alpha)=F^{+}(\alpha)=  \begin{pmatrix}
&B_{11} (\alpha) & B_{12}(\alpha)  & B_{13} (\alpha) & \dots & B_{1K}(\alpha)&\\ 
&B_{21} (\alpha)& B_{22}(\alpha)  & B_{23}(\alpha)  & \dots & B_{2K}(\alpha)&\\ 
&B_{31} (\alpha)& B_{32}  (\alpha)& B_{33}  (\alpha)& \dots & B_{3K}(\alpha)&\\ 
&. & . & . & \dots &  .&\\ 
& .& . & . & \dots    & . &\\ 
&. &.  &  .& \dots  &  .&\\ 
&B_{(N+1) 1}(\alpha) & B_{(N+1) 2} (\alpha) & B_{(N+1) 3} (\alpha) & \dots & B_{(N+1) K}(\alpha)&\\ 
\end{pmatrix}
\end{equation}
\vspace{.5cm}
\noindent The elements of the matrix $P(\alpha)$ are defined as follows:\\
\noindent $\forall \,\,\, i=1: K$ and $ l=1:K$
\begin{equation}
    P_{il}(\alpha)= \mathlarger{\sum}_{m=1}^{N+1} F_{im}(\alpha)\,\,\, B_{ml}(\alpha).
\end{equation}
and its derivative with respect to the fractional differentiation order $\alpha$ can be evaluated as:
\begin{equation}
    \dfrac{\partial P_{il}(\alpha)}{\partial \alpha_j}=\mathlarger{\sum}_{m=1}^{N+1}F_{im}(\alpha) \dfrac{\partial B_{ml}(\alpha)}{\partial \alpha_j}+\mathlarger{\sum}_{m=1}^{N+1} \dfrac{\partial F_{im}(\alpha)}{\partial \alpha_j}B_{ml}(\alpha)
    \label{jac}
\end{equation}
where $\dfrac{\partial B_{ml}(\alpha)}{\partial \alpha_j}$ are the element of the derivative of Moore-Penrose inverse matrix which can be computed using \textbf{Lemma. \ref{lem2}}.  

\noindent $\dfrac{\partial F_{im}(\alpha)}{\partial \alpha_j}$ is derived based on the formulation of (\ref{eq8}) as:\\
\begin{equation}
   \dfrac{\partial F_{im}}{\partial \alpha_j}= \left\{ 
\begin{array}{ll}
-\dfrac{\ln(h)}{h^{\alpha_m}}\left[ \mathlarger{\sum}_{l=1}^{N_1} C_l^{\alpha_m}y(t_{i-l}) + \mathlarger{\sum}_{n=1}^{N_3-1} C_n^{\alpha_m}f_{in}\left[t_i-(N_1-1-n)h\right] \right]-\\ \\  \dfrac{1}{h^{\alpha_m}} \left[\mathlarger{\sum}_{l=0}^{N_1} \dfrac{\partial C_l^{\alpha_m}}{\partial \alpha_m} y(t_{i-l}) + \mathlarger{\sum}_{n=0}^{N_3-1} \dfrac{\partial C_n^{\alpha_m}}{\partial \alpha_m} f_{in}\left[t_i-(N_1-1-n)h\right] \right] & \hspace{1cm} j=m \\ \\
   0 & \hspace{1cm}j \neq m
   \end{array}
   \right.
   \label{F'}
\end{equation}
where $\ln(.)$ is logarithmic function and the partial derivative of the binomial coefficients with respect to $\alpha_j$ is computed iteratively as follows:
\begin{align}
 \dfrac{\partial C_n^{\alpha_j}}{\partial \alpha_j}= \left\{ 
\begin{array}{lll}
    0 \,\,\,\,\,\,\,\,\,\,\,\,\,\,\,\,\,\,\,\,\,\,\,\,\,\,\,\,\,\,\,\,\,\,\,\,\,\,\,\,\,\,\,\,\,\,\,\,\,\,\,\,\,\,\,\,\,\,\,\,\,\,\,\,\,\,\,\,\,\,\,\,\,\,\,\,\,\,\,\,\,\,\,\,\,\,\,\,\,\,\,\,\,\,\,\,\,\,\,\,\,\,\,\,\,\,\,\,\,\,\,\,\,\,\,\,\,\,\,\, n=0 \\ \\
   \dfrac{-1}{n} C_{n-1}^{\alpha_j}+\left(1-\dfrac{1+\alpha_j}{n}\right) \dfrac{\partial C_{n-1}^{\alpha_i}}{\partial \alpha_j} \,\,\,\,\,\,\,\,\, n\geq1
\end{array}
\right.
\end{align}
Substituting (\ref{jac}) in (\ref{J'}) the element of the Jacobian matrix, $J'(\alpha)$ can be computed using the following expression:
\begin{equation}
\label{Jacob'}
    J'_{ij}(\alpha)=\mathlarger{\sum}_{l=1}^{K}\,\,\,\mathlarger{\sum}_{m=1}^{N+1}F_{im}(\alpha)\dfrac{\partial B_{ml}}{\partial \alpha_j}(\alpha) + \mathlarger{\sum}_{l=1}^{K}\dfrac{\partial F_{ij}}{\partial \alpha_j}B_{jl}
\end{equation}
The matrix $J'$ can be expressed as:
\begin{equation}
J'(\alpha)=\begin{pmatrix}
 &\dfrac{\partial J_1}{\partial \alpha_1}& \dfrac{\partial J_1}{\partial \alpha_2}& \dfrac{\partial J_1}{\partial \alpha_3} & \dots &\dfrac{\partial J_1}{\partial \alpha_{N+1}}& \\ 
& . & . & . &\dots & .&\\ 
&. & . & . &\dots & .&\\ 
&.  & . &. &\dots & .&\\ 
& \dfrac{\partial J_K}{\partial \alpha_1}& \dfrac{\partial J_K}{\partial \alpha_2}& \dfrac{\partial J_K}{\partial \alpha_3}& \dots &\dfrac{\partial J_{K}}{\partial \alpha_{N+1}}&
\end{pmatrix}
\end{equation}
\subsection{Proof of convergence}
The iterative character of the above algorithm results from the Newton method which is used to estimate  the fractional differentiation order, $\alpha$. The vector of parameters $p$ is estimated by solving the least square problem. However, we note that this estimate depends on the vector of fractional orders $\alpha$.\\
Thus, we use a standard convergence result of Newton’s method to provide sufficient conditions to ensure the local convergence of the proposed two-stage algorithm. Such a result can be found in Chapter 5 of \cite{kelley1995iterative}. We also make use of two results on Lipschitz functions stating that the product and quotient (denominator not zero) of two bounded Lipschitz functions are Lipschitz.
The following assumptions are needed to prove the convergence result stated in Theorem \ref{theorem}.
\begin{assumption}
\label{assum1}
The input $u(t)$, output $y(t)$ and the initialization function $f_{in}(t)$ are bounded functions in $[0, T]$.
\end{assumption}
%%%
\begin{assumption}
\label{assum2}
A solution $\alpha^*$ to the system $J(\alpha) = 0$ exists.
\end{assumption}
%%%%
\begin{assumption}
\label{assum3}
The Jacobian matrix $J'(\alpha^*)$ is nonsingular.
\end{assumption}
\begin{theorem}
\label{theorem}
Assume that Assumptions $\ref{assum1}-\ref{assum3}$ hold. Then, the proposed proposed two-stage algorithm converges \textit{q-quadratically}.
\end{theorem}
\textbf{Proof.} Based on Theorem 5.1.2, page 71 in \cite{kelley1995iterative}, and given Assumptions \ref{assum2} and \ref{assum3}, it remains to prove that the Jacobian matrix is Lipschitz continuous for $\alpha \in \wp=(n_{i-1},n_i)$ \footnote{$\wp = \prod^{N}_{i=1} \wp_i=\{(a_1,a_2\dots , a_N)| \forall i=1,2,\dots ,N, a_i \in \wp_i\}$ is the generalized Cartesian product of $N$ sets $\wp_1, \wp_2 \dots \wp_N $.} in order to conclude about the convergence of the proposed two-stage algorithm.

The entries of the Jacobian matrix are given by (\ref{Jacob'}) where the terms $F_{im}$, $ \dfrac{\partial F_{im}(\alpha)}{\partial \alpha_j}$, $B_{jl}$ and $\dfrac{\partial B_{ml}(\alpha_j)}{\partial \alpha_j}$ are characterized in (\ref{eq8}), (\ref{F'}), (\ref{B}) and Lemma \ref{lem2}, respectively.

\noindent The norm 1 of the term $J'(\alpha^1) - J'(\alpha^2 )$ is given as follows:
\begin{align}
\label{norm1}
\begin{array}{l}
    \parallel J'(\alpha^2)-J'(\alpha^2) \parallel_1 = \underset{1\leq j\leq N }{\max} \mathlarger{\sum}_{i=1}^{K} \mid J'_{ij}(\alpha^2)-J'_{ij}(\alpha^2) \mid\\ \\ \hspace{3.5cm}
    =\underset{j }{\max} \mathlarger{\sum}_{i=1}^{K} \Bigg |  \left( \mathlarger{\sum}_{l=1}^{K}\,\,\,\mathlarger{\sum}_{m=1}^{N+1}F_{im}(\alpha^1)\dfrac{\partial B_{ml}}{\partial \alpha_j}(\alpha^1) + \mathlarger{\sum}_{l=1}^{K}\dfrac{\partial F_{ij}}{\partial \alpha_j}B_{jl}\right)  \\ \\ \hspace{4cm}-  \mathlarger{\sum}_{l=1}^{K}\,\,\,\mathlarger{\sum}_{m=1}^{N+1}F_{im}(\alpha^2)\dfrac{\partial B_{ml}}{\partial \alpha_j}(\alpha^2) + \mathlarger{\sum}_{l=1}^{K}\dfrac{\partial F_{ij}}{\partial \alpha_j} (\alpha^2)B_{jl} \Bigg | \\ \\ \hspace{3.5cm}
    =\mathlarger{\sum}_{i=1}^{K} \Bigg |   \mathlarger{\sum}_{l=1}^{K}\,\,\,\mathlarger{\sum}_{m=1}^{N+1} \left\{ F_{im}(\alpha^1)\dfrac{\partial B_{ml}}{\partial \alpha_{j^*}}(\alpha^1) - F_{im}(\alpha^2)\dfrac{\partial B_{ml}}{\partial \alpha_{j^*}}(\alpha^2) \right\}+ \\ \\ \hspace{4cm} \left( \mathlarger{\sum}_{l=1}^{K}\left\{\dfrac{\partial F_{ij^*}}{\partial \alpha_{j^*}} (\alpha^1)B_{j^*l}(\alpha^1)-\dfrac{\partial F_{ij^*}}{\partial \alpha_{j^*}} (\alpha^2)B_{j^{*}l}(\alpha^2)\right\}\right)\Bigg |
    \end{array}
\end{align}
where $j^*$ is the column index corresponding to the maximum of the sum of the row components. Then, using the triangle inequality, (\ref{norm1}) can be written as follows:
\begin{equation}
\label{ineq}
\begin{array}{ll}
\parallel J'(\alpha^2)-J'(\alpha^2) \parallel_1 \,\, \leqslant \\\\
\mathlarger{\sum}_{i=1}^{K} \,\,\, \mathlarger{\sum}_{l=1}^{K}\,\,\,\mathlarger{\sum}_{m=1}^{N+1} \left | F_{im}(\alpha^1)\dfrac{\partial B_{ml}}{\partial \alpha_{j^*}}(\alpha^1) - F_{im}(\alpha^2)\dfrac{\partial B_{ml}}{\partial \alpha_{j^*}}(\alpha^2) \right | \\\\ \hspace{.3cm}+ \mathlarger{\sum}_{i=1}^{K} \,\,\,  \mathlarger{\sum}_{l=1}^{K} \left| \dfrac{\partial F_{ij^*}}{\partial \alpha_{j^*}} (\alpha^1)B_{j^{*}l}(\alpha^1)-\dfrac{\partial F_{ij^*}}{\partial \alpha_{j^*}} (\alpha^2)B_{j^*l}(\alpha^2) \right|
\end{array}
\end{equation}
Given the right hand side of inequality (\ref{ineq}) and using properties on Lipschitz functions, if $F_{im}(\alpha)$ and $\dfrac{\partial F_{ij}}{\partial \alpha_j}(\alpha)$ are Lipschitz continuous and  $B_{jl}(\alpha)$ and $\dfrac{\partial B_{ml}}{\partial \alpha_j}(\alpha)$ are bounded then we can conclude about the  Lipschitz continuity of $J'(\alpha)$.\\

\noindent Given that $C_l^\alpha$ and $\dfrac{\partial C_l^\alpha}{\partial \alpha}$ is high-order polynomial of $\alpha$ then they are Lipschitz continuous. In addition, $\left(\dfrac{1}{h^\alpha}\right)$ is bounded on bounded local interval of $\alpha$. Accordingly, and taking into account \textit{ Assumption \ref{assum1}} we can insure the Lipschitz continuity of $F_{im}(\alpha)$ and $\dfrac{\partial F_{ij}}{\partial \alpha_j}(\alpha)$.\\

\noindent With respect to the boundedness of $B_{jl}(\alpha)$ and $\dfrac{\partial B_{ml}}{\partial \alpha_j}(\alpha)$, based on the proof in [] %\ref{}%
that confirm that if $F(\alpha)$ has a constant rank for $\alpha^*-\delta<\alpha\leq \alpha^*+\delta$ then its Moore-Penrose inverse $B$ is differentiable at $\alpha^*$ and its derivative is computed using \textit{Lemma \ref{lem2}.} Accordingly the of $B_{jl}$ and its derivative follows from this condition.
%%%%%%%%%%%%%%%%%%%%%%%%%%%%%%%%%%%%%%%%%%%%%%%%%%%%%%%%%%%%%%%%%%%%%%%%%%%%%%%%%%%%%%%%
\section{Numerical Results}
The performance of the proposed method is illustrated through different numerical examples. Furthermore, potential applications of the algorithm are presented, which consists of estimating parameters and fractional differentiation orders of 1) fractional-order two-element arterial Windkessel (\textit{F-WK2}) model and 2) fractional-order neurovascular model. In the following analysis of the results, we consider the percent relative error $Re(\%)$ as a measure of estimation accuracy. In the case of parameter $p$ estimation, the $Re_p(\%)$ is defined as:
\begin{equation}
    Re_p=\dfrac{|\hat{p}-p|}{|p|}\times 100,
\end{equation}
where $\hat p$ denotes the estimate of $p$. In the case of signal estimation $y$, the $Re_y(\%)$ is defined as:
\begin{equation}
    Re_y=\dfrac{||\hat{y}-y||_2}{||y||_2}\times 100,
\end{equation}
where $\hat y$ denotes the estimate of the real measurement $y$. 
\subsection{Example 1}
\begin{figure}[!t]
        \centering
        \includegraphics[width=.75\linewidth]{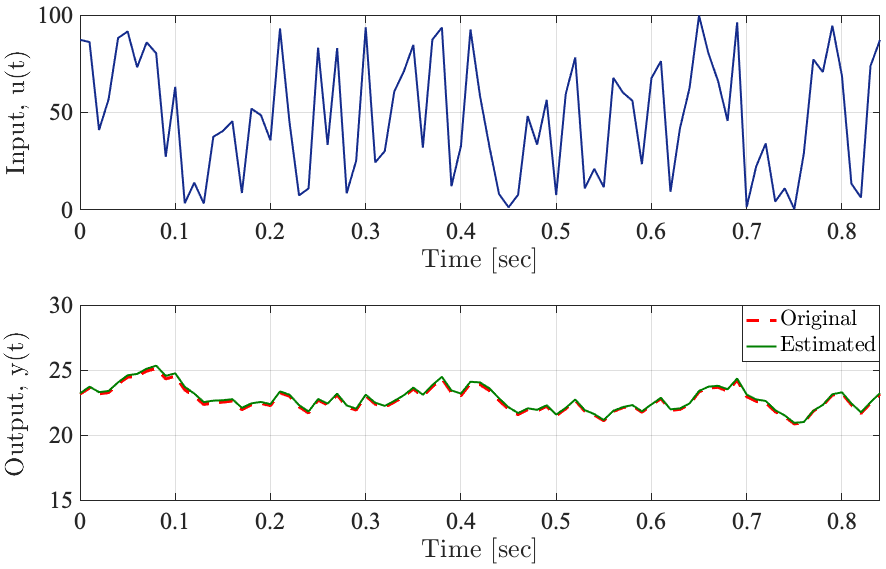}
\caption{\textcolor{black}{Estimated output signal along with the real one using random input signal. Here we use $N_0=15$ and $N_C=10$. The sampling rate $h=0.01$ has been used in this simulation.}}
\label{fig5_1}
\end{figure}
We consider the following fractional-order system:
\begin{equation}
\label{numerical_example}
    y(t)+a_1\,\,\,_{t_{in}}D_t^\alpha y(t)=b\,\,\, u(t) \, \, \, \, \, \,  t \in [0,\,\,\, 0.84]
\end{equation}
\textcolor{black}{Here we consider two numerical cases based on the nature of the input signal $u(t)$, namely: Random input signal and periodic pulse train input signal.
\subsubsection{Random input signal}
In this example, we consider the system described in (\ref{numerical_example}) with a random input signal whose amplitude varies between $0$ and $100$. The parameters and fractional differentiation order are $a_1=1$, $b=0.5$ and $\alpha=0.7$. A sensitivity analysis of the proposed algorithm with respect to the length of the time-varying history function-based output fractional-order derivative initialization, $N_3$, and the length of the original signal was performed. In addition, a sensitivity analysis with respect to the initial guess of Newton’s method (the initial guess of the fractional differentiation order $\alpha^0$) is performed.
%%%%%%%%%%%%%%%%%%%%%%%%%%%%%%%%%%%%%%%%%%%%%%%%%%%%%%%%%%%%%%%%%%%%%%%%%%%%%%%%%%%%%%%%%%
%%%%
\begin{figure}[!t]
        \centering
        \includegraphics[width=.75\linewidth]{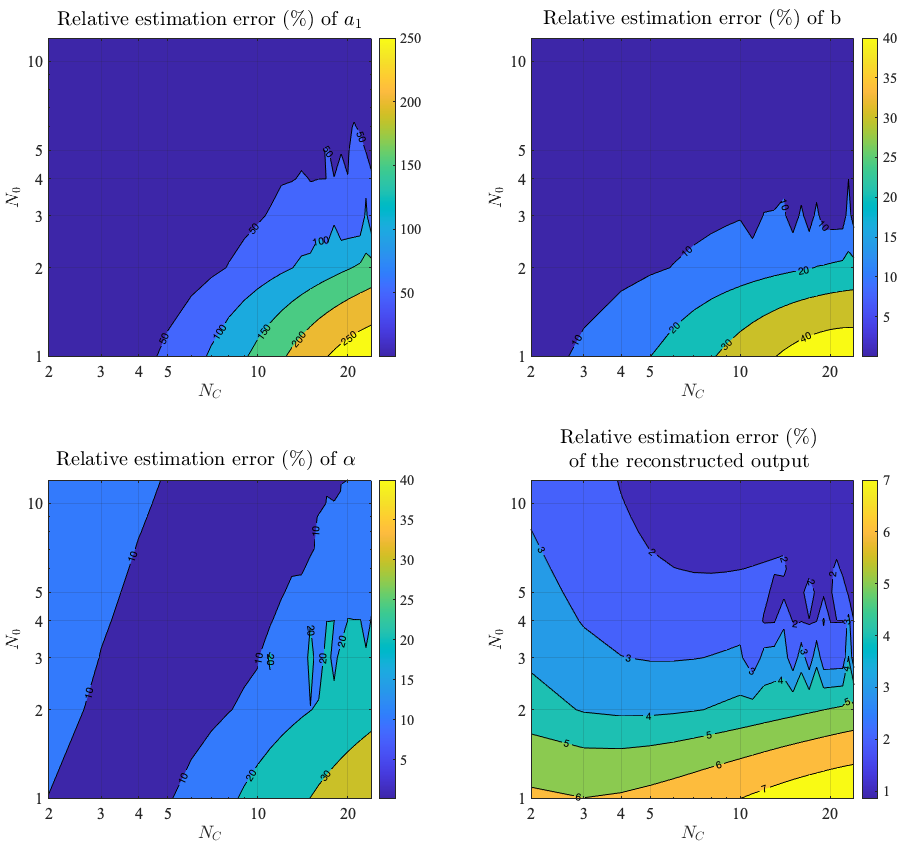}
\caption{\textcolor{black}{Contour plot of the relative error (\%) with respect to the number of cycles based time-varying history function, $N_C$ and the number of original cycles based for estimation, $N_0$ for: parameter ($a_1$), parameter ($b$), the fractional differentiation order ($\alpha$) and the reconstructed output signal ($y$). The sampling rate $h=0.01$ has been used in this simulation.}}
\label{fig41}
\end{figure}
%%%%
\begin{figure}[!t]
        \centering
        \includegraphics[width=.85\linewidth]{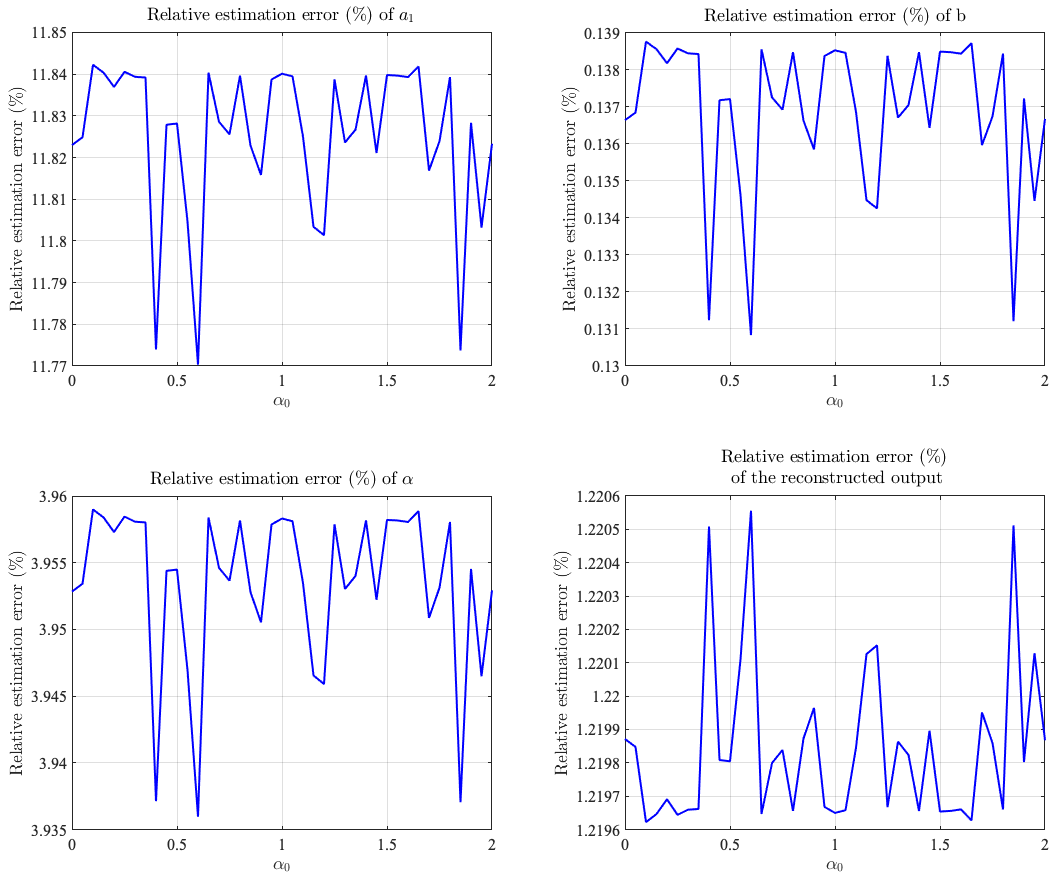}
\caption{\textcolor{black}{Relative error (\%) with respect to the initial guess of the fractional differentiation order $\alpha_0$ for: parameter ($a_1$), parameter ($b$), the fractional differentiation order ($\alpha$) and the reconstructed output signal ($y$).The sampling rate $h=0.01$ has been used in this simulation.}}
\label{fig_alpha_0_random}
\end{figure}
%%%%%%%%%%%%%%%%%%%%%%%%%%%%%%%%%%%%%%%%%%%%%%%%%%%%%%%%%%%%%%%%%%%%%%%%%%%%%%%%%%%%%%%%%%
We note $N_C=N_3/L$ and $N_0=K/L$ where $L$ corresponds to the length of the original output signal and $K$ is the total number of samples used for estimation, following the previous notations, it is defined as $K=N_1-N_3$.\\
%%%%%%%%%%%%%%%%%%%%%%%%%%%%%%%%%%%%%%%%%%%%%%%%%%%%%%%%%%%%%%%%%%%%%%%%%%%%%%%%%%%%%%%%%%
\noindent Figure \ref{fig5_1} shows the input signal and the estimated output signal along with the original one for $N_0=15$ and $N_C=10$. In this case, the relative errors of the parameters estimate $a_1$ and $b$ were $11.77 \%$ and $ 0.13 \%$ respectively, whereas, for the fractional differentiation order estimate, it was $3.93\%$ and the reconstructed output around $1.22\%$. \textcolor{black}{It is clear from these results that we have a good estimation performance for the output signal. Although the estimation error of the parameters and the fractional differentiation order is not the best, we still have a good reconstruction of the output. The reason is that in this case, we cannot insure the global identifiability of the system (\ref{numerical_example}).} \\
%%%
\noindent Figure \ref{fig41} shows a contour plot of the relative errors (\%) with respect to the number of cycles based time-varying history function, $N_C$, and the number of original cycles based for estimation, $N_0$ for the parameters, fractional differentiation order and the reconstructed output. Overall, from these subplots it is clear that increasing both $N_C$ and $N_0$ leads to a decrease in the relative error and so better estimation for the parameters, fractional differentiation order, and the reconstructed output. For a fixed $N_C$ we noticed that as $N_0$ increases the relative error decreases.\\
With regards to the unknown parameters and $\alpha$, there are optimal values of ($N_C^*, N_0^*$) that lead to the minimum error. Any combination larger than  ($N_C^*, N_0^*$) might lead to a small increase in the error. With respect to the output relative error, we noticed that increasing more $N_0$ leads to better performances. For small values of $N_0=1,2\dots 5$ it is clear that there is the optimal combination ($N_C^*, N_0^*$) that leads to the best performance.\\
\noindent Figure \ref{fig_alpha_0_random} summarizes the results of the relative errors of the estimated unknowns when the initial guess of Newton’s method is varying. In this simulation we varied $\alpha^0$ from $0.05$ to $2$ with a step of $0.05$. It is clear that the proposed algorithm is not very sensitive to initial guesses in this case.
\begin{figure}[!t]
        \centering
        \includegraphics[width=.70\linewidth]{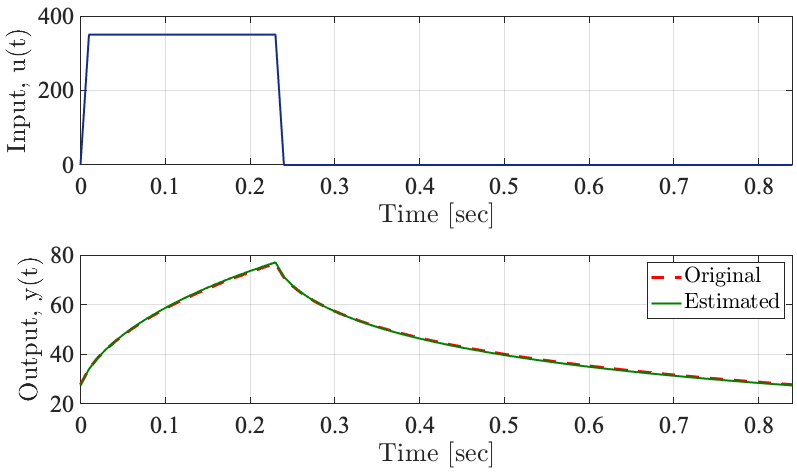}
\caption{\textcolor{black}{Estimated output signal along with the real one using random input signal. Here we use $N_0=15$ and $N_C=10$. The sampling rate $h=0.01$ has been used in this simulation.}}
\label{fig_reconstruction_anlytic}
\end{figure}
\begin{figure}[!t]
        \centering
        \includegraphics[width=.75\linewidth]{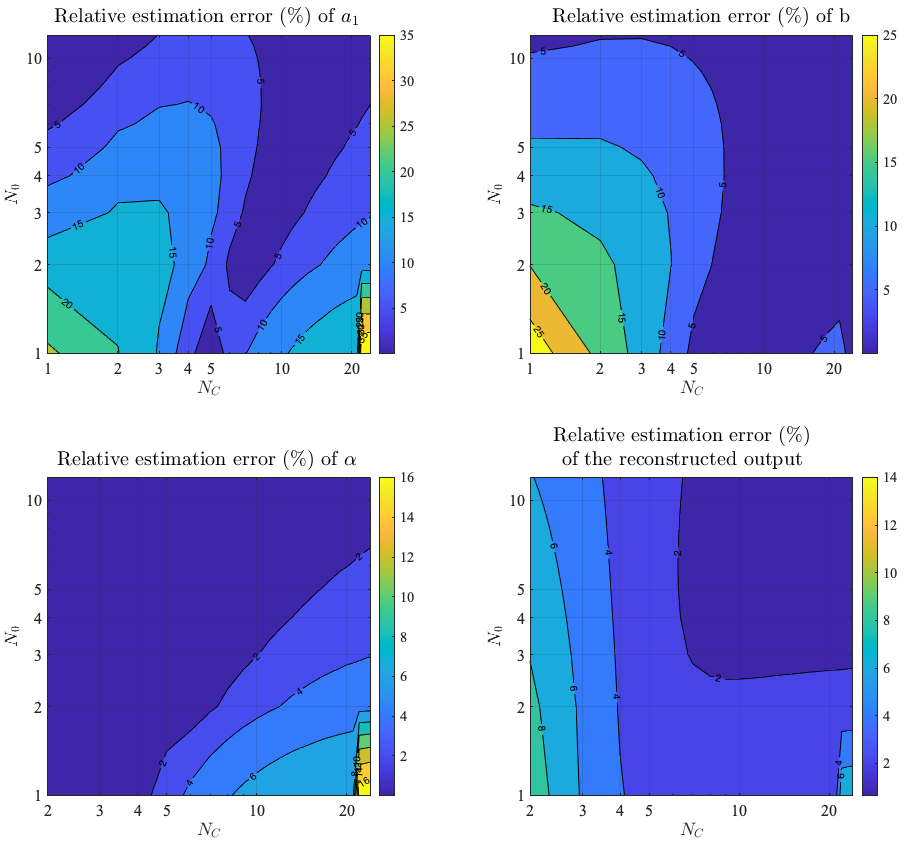}
\caption{\textcolor{black}{Contour plot of the relative error (\%) with respect to the number of cycles based time-varying history function, $N_C$ and the number of original cycles based for estimation, $N_0$ for: parameter ($a_1$), parameter ($b$), the fractional differentiation order ($\alpha$) and the reconstructed output signal ($y$). The sampling rate $h=0.01$ has been used in this simulation.}}
\label{fig52}
\end{figure}
%%%%%%%%%%%%%%%%%%%%%%%%%%%%%%%%%%%%%%%%%%%%%%%%%%%%%%%%%%%%%%%%%%%%%%%%%%%%%%%%%%%%
\begin{figure}[!t]
        \centering
        \includegraphics[width=.75\linewidth]{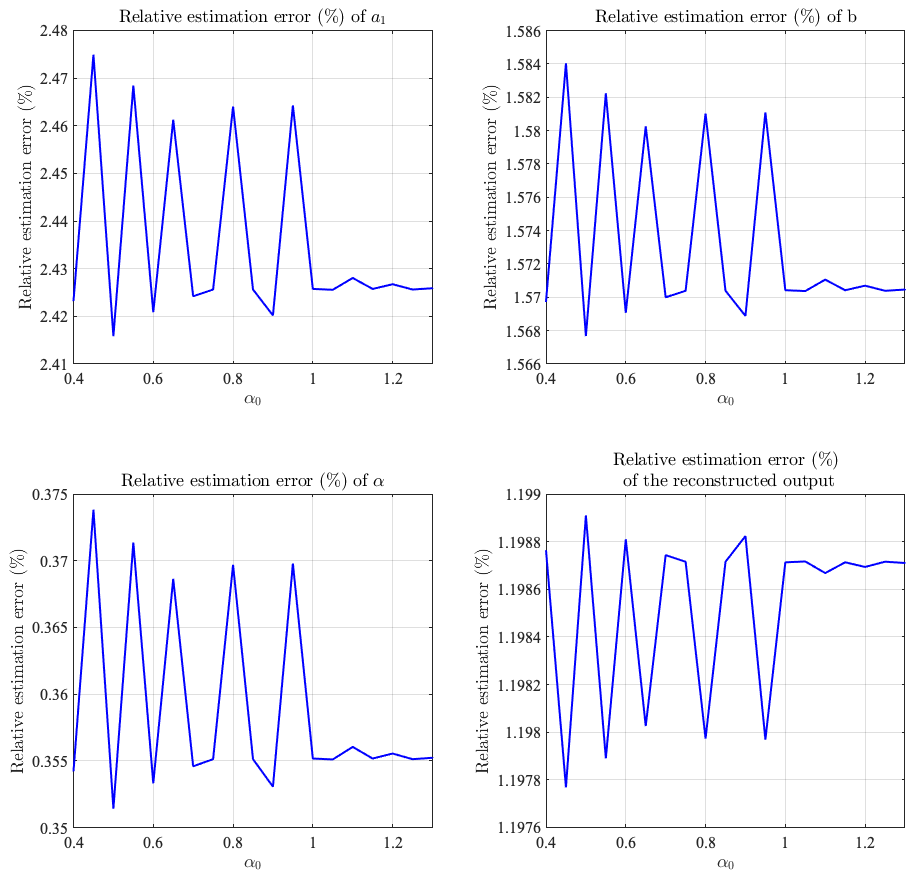}
        \vspace{-.75cm}
\caption{\textcolor{black}{Relative error (\%) with respect to the initial guess of the fractional differentiation order $\alpha_0$ for: parameter ($a_1$), parameter ($b$), the fractional differentiation order ($\alpha$) and the reconstructed output signal ($y$). The sampling rate $h=0.01$ has been used in this simulation.}}
\label{fig_alpha_0_analytic}
\end{figure}
%%%%%%%%%%%%%%%%%%%%%%%%%%%%%%%%%%%%%%%%%%%%%%%%%%%%%%%%%%%%%%%%%%%%%%%%%%%%%%%%%%%%
\subsubsection{Periodic pulse train input signal}
In this example we consider the system described in (\ref{numerical_example})  where the parameters and fractional differentiation order are $a_1=1$, $b=0.5$ and $\alpha=0.7$. The input $u(t)$ is taken as a periodic pulse train of duration $T= 0.84 \,\,\,[s]$, an amplitude $A=350$ and a duty cycle ($d_C=0.27$). The time step used in this example is $h=0.01$.}
\textcolor{black}{Similar to the previous example, a sensitivity analysis of the proposed algorithm with respect to the length of the time-varying history function-based output fractional-order derivative initialization, the length of the original signal and the initial guess of Newton’s method is performed.
Figure \ref{fig_reconstruction_anlytic} summarizes the reconstruction results of the output of the fractional-order system (\ref{numerical_example}) using a periodic pulse train signal as an output. In this simulation we used ($N_0= 15; N_C= 10$). The relative errors of this reconstruction is $1.19 \%$, for the unknown parameter $a_1$ is $2.42 \%$ and $b$ is $1.56\%$ and the fractional-differentiation order is $0.35\%$.\\
%%%
\noindent In figure \ref{fig52}, we summarize the results of the relative errors of the estimated unknowns and the reconstructed output for a different combination of ($N_0;N_C$). \\ 
Figure \ref{fig_alpha_0_analytic} summarizes the results of the relative errors of the estimated unknowns when the initial guess of the fractional differentiation order, $\alpha^0$, is varying from $0.4$ to $1.25$ with a step $0.05$.}
%%%%%%%%%%%%%%%%%%%%%%%%%%%%%%%%%%%%%%%%%%%%%%%%%%%%%%%%%%%%%%%%%%%%%%%%%%%%%%%%%%%%%%%%%%
\subsubsection{Joint estimation of the parameters and the fractional differentiation order of F-WK2}
\begin{figure}[!t]
    \centering
        \includegraphics[width=.5\linewidth]{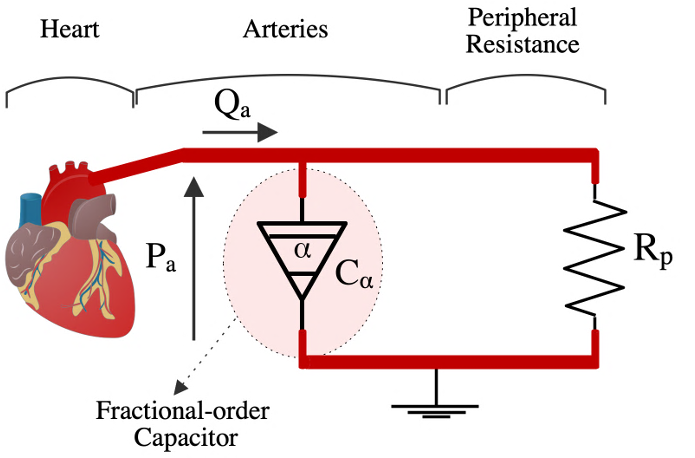}
\caption{Fractional-order tow-element arterial Windkessel circuit model. $C_{\alpha}$ represents the arterial compliance and $R_p$ represents the peripheral resistance.}
\label{fig53}
\end{figure}

A potential application of the proposed algorithm is to estimate the hemodynamic parameters and the differentiation order of the fractional-order two-element arterial Windkessel (\textit{F-WK2}). As shown in Fig. \ref{fig53}, \textit{F-WK2} represents the heart as a current source that pumps blood into the arterial system, which is lumped into a single fractional-order capacitor and a single resistance \cite{bahloul2020assessment,laleg2021mathematical}. The fractional-order capacitor, characterized by two parameters ($C_\alpha, \alpha$), represents the apparent arterial compliance by taking into account the viscoelastic properties of the vascular network via the fractional-order factor \cite{bahloul2022human,bahloul2021towards}. The resistance $R_p$ accounts for the total peripheral arterial resistance.
The \textit{F-WK2} system is given as follows:
\begin{equation}
    P_a(t)+\tau_\alpha\,\,\,_{t_{in}}D_t^\alpha P_a(t)=R_p\,\,\, Q_a(t), \,\,\, t\in [0, T_C]
    \label{eqfwk2}
\end{equation}
where $\small \tau_\alpha=C_\alpha \cdot R_p$, $\small 0\leq\alpha\leq1$ and $\small T_C$ is the cardiac period. $ \small P_a$ corresponds to the aortic blood pressure, and $\small Q_a$ denotes the aortic blood flow.
\begin{figure}[!t]
    \centering
\includegraphics[width=.65\linewidth]{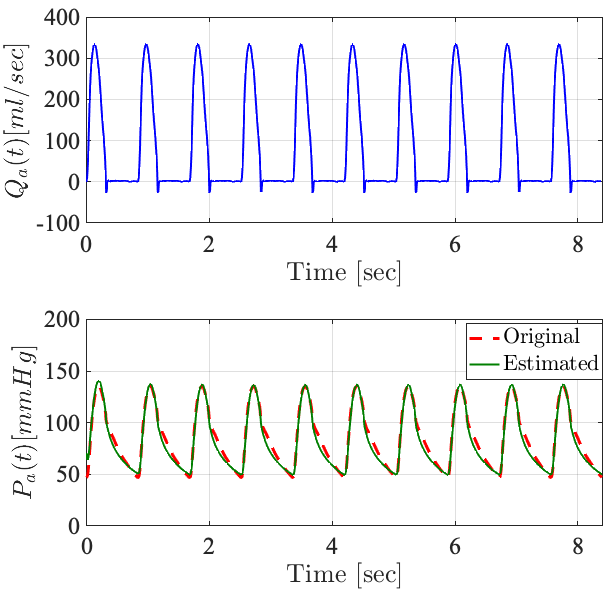}
\caption{Estimated aortic blood pressure using \textit{F-WK2} along with the real one. Here we use $N_0=10$ and $N_C=25$. The relative errors of the reconstructed output is $5.22 \%$, $\tau_\alpha=1.15$ and $R_p=1.13$. The sampling rate $h=0.01$ has been used in this simulation.}
\label{fig54}
\end{figure}
%%%%%%%%%%%%%%%%%%%%%
%%%%%%%%%%%%%%%%%%%%%%%%%%%%%%%%%%%%%%%%%%%%%%%%%%%%%%%%%%%%%%%%%%%%%%
%%%%%%%%%%%%%%%%%%%%%%%%
\noindent To fully identify \textit{F-WK2}, the hemodynamic parameters and the fractional differentiation order have to be estimated from measured flow and pressure waves from accessible arterial locations. This is also known as the hemodynamic inverse problem \cite{bernhard2012non}. The proposed algorithm has been applied to an \textit{in-silico} data generated from one-dimensional model \cite{charlton2019modeling}. Here, the aortic blood pressure is considered the output of the fractional-order system, while the aortic blood flow is considered the input. To initialize the fractional-order derivative of the output ($D_t^\alpha P_a(t)$), we used a certain number of cycles, $N_C$ of the same output signal as a time-varying history function. We denote by $N_0$ the number of original cardiac cycles used for estimation.
A sensitivity analysis of the applied algorithm with respect to $N_C$ and ($N_0$) is performed.\\
\noindent \textcolor{black}{\noindent Fig. \ref{fig54} summarizes the reconstruction results of $10$ cardiac cycles of the aortic blood pressure using a real input and the estimated parameters and the fractional differentiation order. This reconstruction used $25$ cycles to initialize the fractional-order derivative of the output and the relative error was around $ 5.22\%$. The estimate of the constant time is $\hat \tau_\alpha=1.15$ and  the peripheral resistance $\hat R_p=1.13$.\\
\noindent Figure \ref{fig55} summarizes the sensitivity analysis of the relative error of the reconstructed aortic blood pressure (output of the system (\ref{eqfwk2})) with respect $N_C$ and $N_0$. It is clear that as we increase $N_C$ the relative error of the estimation decreases. Figure \ref{fig_alpha_0_real} summarizes the results of the relative errors of the estimated output the initial guess is varying from $0.35$ to $1.2$ with a step of $0.05$} 
\begin{figure}[!t]
\begin{minipage}[c]{0.49\linewidth}
    \centering
\includegraphics[width=.85\linewidth]{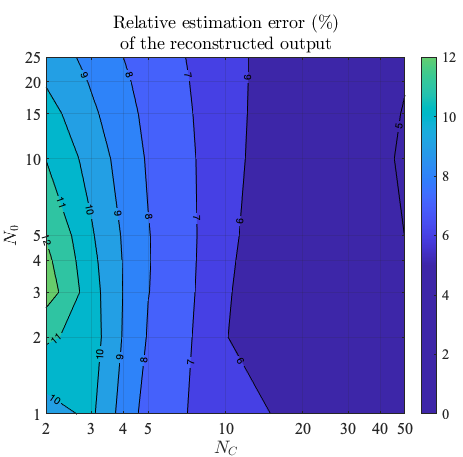}
\caption{Relative estimation error vs. number of cycles based time-varying history function, $N_C$ for one cycle output aortic blood pressure signal ($N_0=1$). The sampling rate $h=0.01$ has been used in this simulation.}
\label{fig55}
\end{minipage}
\hspace{0.05cm}
\begin{minipage}[c]{0.49\linewidth} 
    \centering
  \includegraphics[width=.95\linewidth]{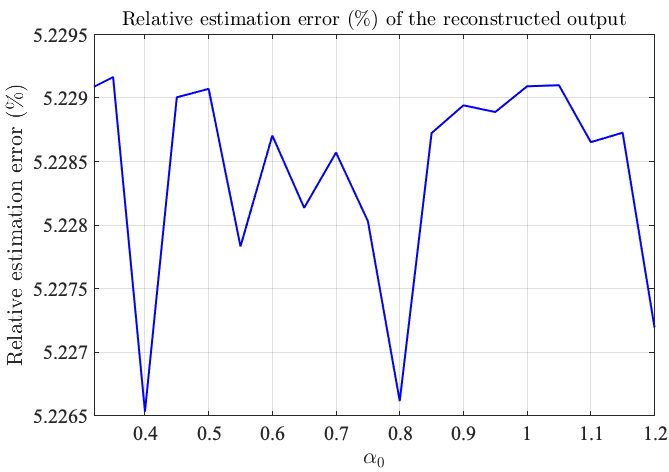}
\caption{\textcolor{black}{Relative error (\%) with respect to the initial guess of the fractional differentiation order $\alpha_0$ for the reconstructed output signal, the aortic blood pressure. The sampling rate $h=0.01$ has been used in this simulation.}} 
\label{fig_alpha_0_real}
\end{minipage}        
\end{figure} 
%%%%%%%%%%%%%%%%%%%%%%%%%%%%%%%%%%%%%%%%%%%%%%%%%%%%%%%%%%%%%%%%%%%%%%%%%%%%%%%%%%%%%%%%%
\subsection{Example 2}
We consider the following example:
\begin{equation}
    y(t)+a_1\, D^{\alpha_1}_t y(t)+ a_2 \,D^{\alpha_2}_t y(t)= u(t), \,\,\, t\in [0, 10]
    \label{example2}
\end{equation}
where the parameters and fractional differentiation orders are $a_1=3$, $a_2=2$ and $\alpha_1=1.5$, $\alpha_2=0.5$. Here we consider two numerical cases based on the nature of the input signal $u(t)$, namely: The $sinc$ function as the input signal and the square input signal.
%%%%%%%%%%%%%%%%%%%%%%%%%%%%%%%%%%%%%%%%%%%%%%%%%%%%%%%%%%%%%%%%%%%%%%%%%%%%%%%%%%%%
\subsubsection{$ Sinc$ function as an input signal}
%%%%%%%%%%%%%%%%%%%%%%%%%%%%%%
\begin{figure}[!t]
        \centering
        \includegraphics[width=.75\linewidth]{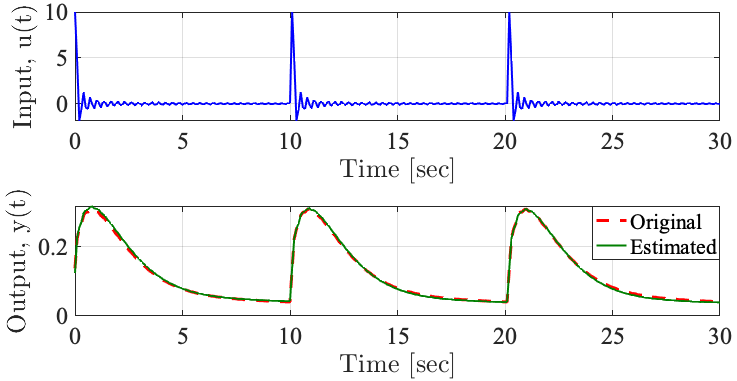}
\caption{\textcolor{black}{Estimated output signal along with the real one using random input signal. Here we use $N_0=3$ and $N_C=10$. The sampling rate $h=0.1$ has been used in this simulation.}}
\label{yy_sinc}
\end{figure}
%%%%%%%%%%%%
\begin{figure}[!t]
        \centering
        \includegraphics[width=.9\linewidth]{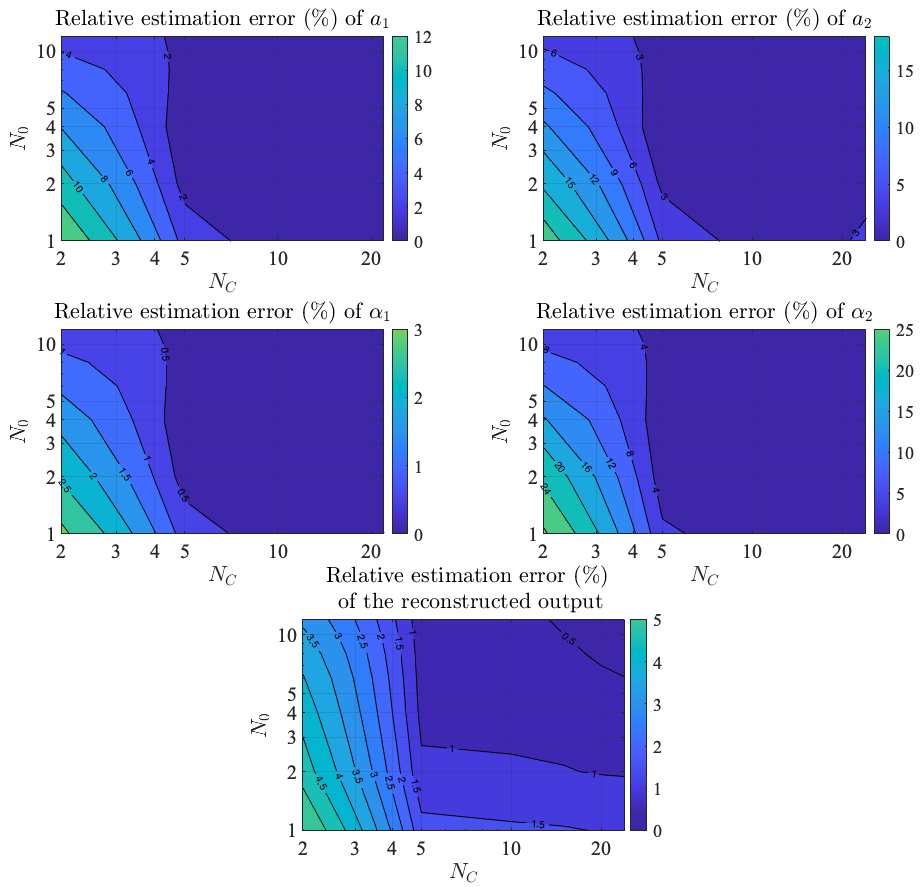}
\caption{\textcolor{black}{Contour plot of the percent relative error (\%) with respect to $N_C$ and $N_0$ for parameters, fractional differentiation orders and the reconstructed output. The sampling rate $h=0.1$ has been used in this simulation.}}
\label{param_neuro_sinc}
\end{figure}
%%%%%%%%%%
In this case, we consider the following profile for the input $u(t)$ in the system (\ref{example2}):
\begin{equation}
    u(t)= 10 \sinc(2 \pi t)
\end{equation}
Figure \ref{yy_sinc} shows the reconstruction result of the estimated output using $N_C=10$ and $N_0=3$. The relative errors of the unknown parameters ($a_1,a_2,\alpha_1,\alpha_2$) are ($1.45\%,1.60\%,0.33\%,3.61\%$) respectively and for the reconstructed output is $0.88\%$.\\
\noindent In figures \ref{param_neuro_sinc}, we summarize the results of the relative errors of the estimated unknowns parameters, fractional differentiation orders, and the reconstructed output for a different combination of ($N_0;N_C$). From these results, it is clear that as the $N_C$ increases the estimation accuracy increases. It is worth noting that for a small value of $N_C$ increasing $N_0$ enhances the estimation accuracy.
%%%%%
\subsubsection{Square input signal}
\begin{figure}[!t]
        \centering
        \includegraphics[width=.75\linewidth]{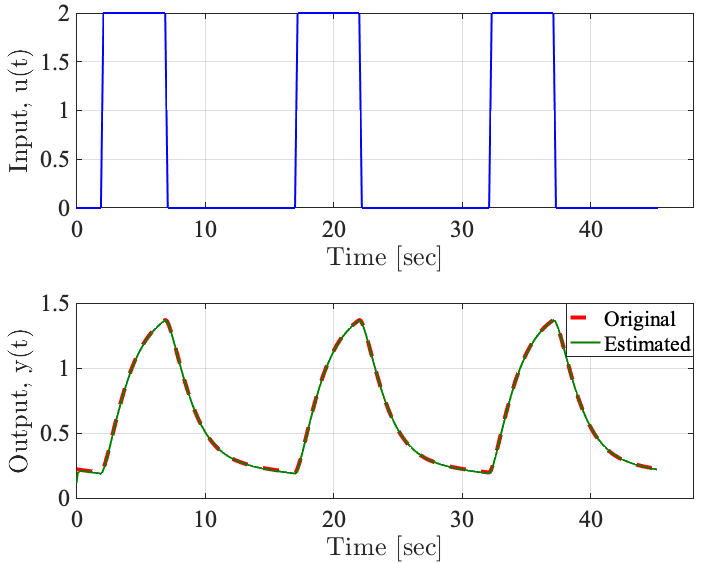}
\caption{\textcolor{black}{Estimated output signal along with the real one using random input signal. Here we use $N_0=3$ and $N_C=15$.}}
\label{fig_neuro_signal_analytic}
\end{figure}
\begin{figure}[!t]
        \centering
        \includegraphics[width=1\linewidth]{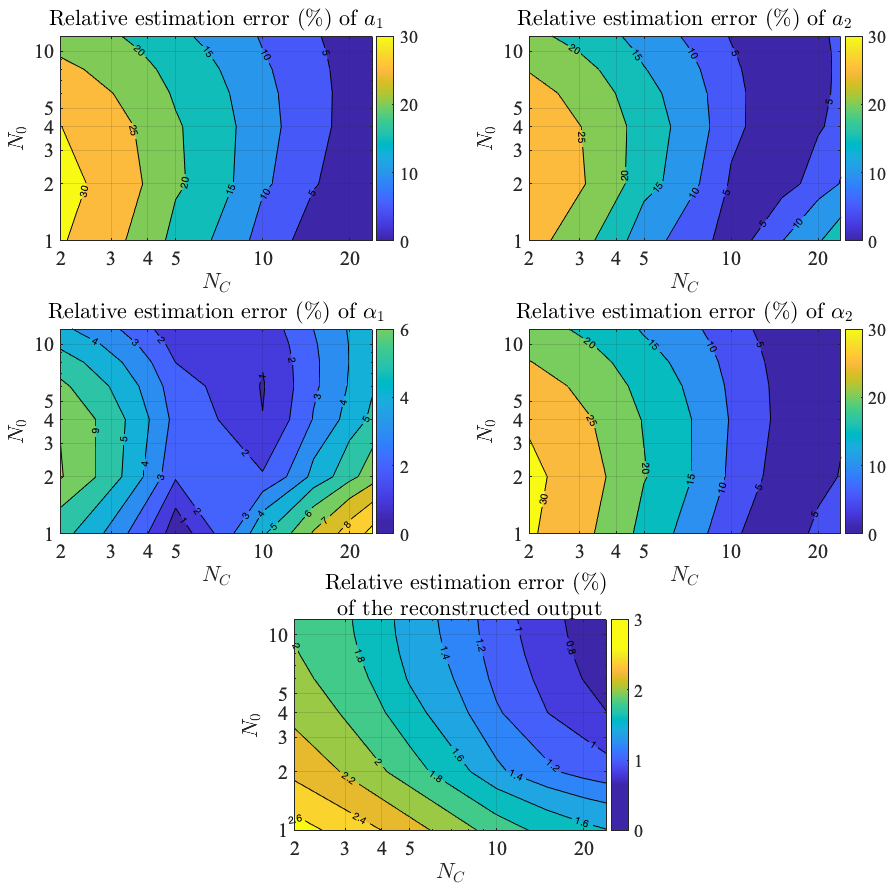}
\caption{\textcolor{black}{Contour plot of the relative error (\%) with respect to  $N_C$ and $N_0$ for parameters, fractional differentiation orders and the reconstructed output signal. The sampling rate $h=0.01$ has been used in this simulation.}}
\label{param_neuro}
\end{figure}
In this example the input $u(t)$ of the fractional-order system (\ref{example2}) is taken as a square function activated at $t=2$ and deactivated at $t=7$. Figure \ref{fig_neuro_signal_analytic} shows the reconstruction result of the estimated output using $N_C=N_0=3$. 

\noindent The relative errors of the unknown parameters ($a_1,a_2,\alpha_1,\alpha_2$) are (4.19\%, 4.17\%, 1.07\%, 5.84\%) respectively and for the reconstructed output is $0.86\%$. It is clear in this example that although the percent relative error of the reconstructed output is very low, the relative errors of the unknown parameters and fractional differentiation estimates are higher. This can be explained by the fact that fractional-order systems with more than one fractional differentiation order are not structurally globally identifiable \cite{alavi2015structural}. Figure \ref{param_neuro} shows the percent relative errors of the estimation of the parameters ($a_1;a_2$), the fractional differentiation orders ($\alpha_1;\alpha_2$) and the reconstructed output $y(t)$. Based on these results, overall, we have a good reconstruction. In all cases, the percent relative error does not exceed $3\%$. The best performance is achieved with large values of $N_C$ and $N_0$. Concerning the parameters and fractional differentiation orders, we obtained the same aspect. However, the error is larger than the ones obtained in the reconstruction of the output. 

\subsubsection{Joint estimation of the parameters and the fractional differentiation order for a fractional neurovascular mode}
\begin{figure}[!t]
        \centering
        \includegraphics[width=.9\linewidth]{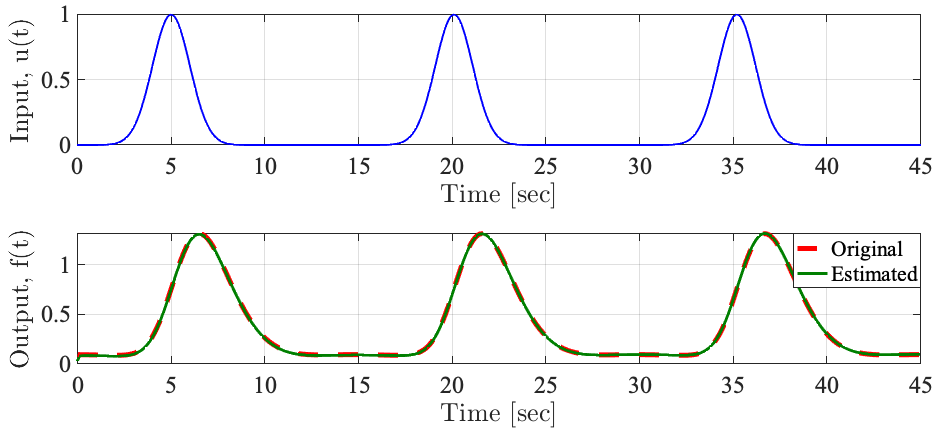}
\caption{\textcolor{black}{Estimated output signal along with the real one using random input signal. Here we use $N_0=3$ and $N_C=10$.}}
\label{fig_neuro_syn}
\end{figure}
A potential application of the proposed method is to estimate the neural activity and the fractional differentiation orders from the Cerebral Blood Flow (CBF) measurements. In \cite{belkhatir2014fractional}, a fractional model for the neurovascular coupling in the brain has shown a better characterization of the cerebral hemodynamic response. This model relates the neural activity, considered as input, to the CBF considered as output. To fully identify the fractional neurovascular model, the input and the fractional differentiation orders have to be estimated from CBF data. The nature of the neural activity can be of two types: the block design paradigm, which corresponds to tasks lasting for extended periods of time, and the event-related paradigm, which corresponds to tasks lasting for shorter periods. 
This section focuses on estimating the event-related type of neural activity usually modeled by Gaussian functions.
The neurovascular model is given as follows:
\begin{equation}
    D^{\alpha_1}_t f(t)+ k \,D^{\alpha_2}_t f(t) + \gamma f(t)=  u(t), \,\,\, t\in [0, 15]
    \label{neuro}
\end{equation}
where $1 \leq \alpha_1 \leq 2$, $0 \leq \alpha_2 \leq 1$ are the fractional differentiation orders, $f(t)$ is the temporal CBF and $u(t)$ is the temporal neuronal activity. The constants $k$ and $\gamma$ are the flow signal decay rate and the flow-dependant elimination constant, respectively. Their values are given in as: $k=0.65$, $\gamma=0.41$. 
\begin{figure}[!t]
        \centering
        \includegraphics[width=.95\linewidth]{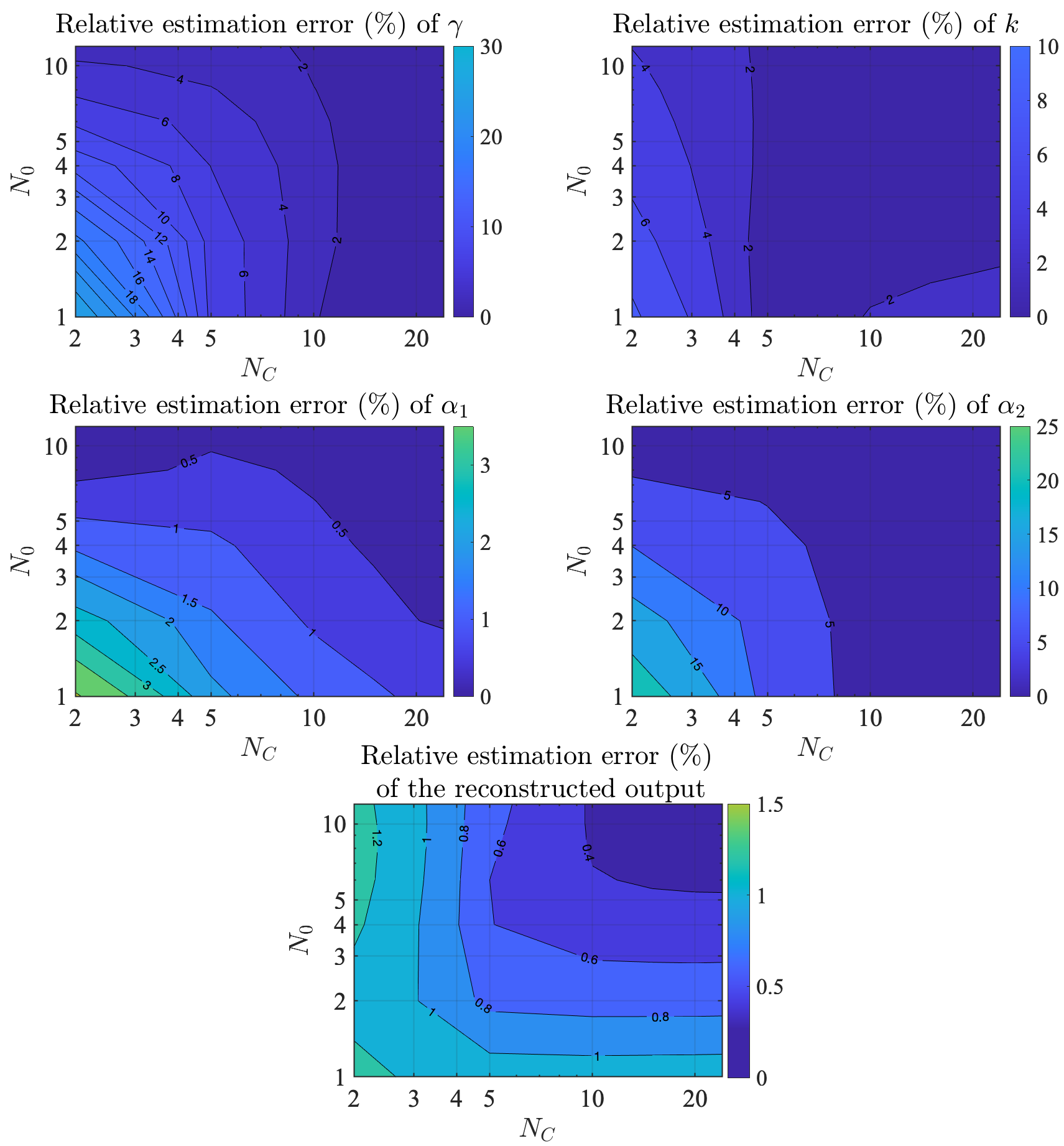}
\caption{\textcolor{black}{Contour plot of the relative error (\%) with respect to  $N_C$ and $N_0$ for parameters, fractional differentiation orders and the reconstructed output signal. The sampling rate $h=0.1$ has been used in this simulation.}}\label{FFF}
\end{figure}
\noindent The tested input profile is Gaussian which can be expressed as:\\
\begin{equation}
  u(t)=\exp \left[-(t-5)^2 \right]  
\end{equation}
\noindent We use a synthetic data set which have been generated from discretizing the neurovascular model (\ref{neuro}) and fixing ${\alpha_1=1.7, \alpha_2=0.6}$. 

\noindent Figure \ref{fig_neuro_syn} shows the reconstruction result of the estimated output using $N_C=10$ and $N_0=3$. The relative errors of the unknown parameters ($k$,$\gamma$,$\alpha_1,\alpha_2$) are ($1.32\%,1.64\%,0.62\%,1.67\%$) respectively and for the reconstructed output is $0.57\%$.\\
\noindent In figures \ref{FFF}, we summarize the results of the relative errors of the estimated unknown parameters, fractional differentiation orders, and the reconstructed output for a different combination of ($N_0; N_C$). From these results, it is clear that as the $N_C$ increases the estimation accuracy increases. It is worth noting that for a small value of $N_C$ increasing $N_0$ enhances the estimation accuracy.
\section{Conclusion}
The concept of initialized fractional calculus is of great pertinence for the uniqueness of solutions to fractional forward problems. This report investigates their significance for solving inverse problems. We proposed an initialized estimation algorithm to jointly estimate the parameters and the fractional differentiation orders while designing an output-dependent history function based-initialization. After reformulating the estimation problem in the discrete space, we proposed an algorithm of \textcolor{black}{two} steps that study the effect of the length of the initialization function while estimating the unknown parameters. We applied the pipeline to different numerical examples. Furthermore, potential applications of the proposed algorithm are presented, which consists of joint estimation of  parameters and fractional differentiation orders of a fractional-order arterial Windkessel and neurovascular models. In future work, we will propose a multi-stage estimation algorithm based on modulating functions that are known to be robust to noise.
\bibliographystyle{IEEEtran}
\bibliography{references}
\end{document}